\documentclass[showpacs,showkeys,amsmath,amssymb]{revtex4}
\usepackage{amsmath}
\usepackage{amsfonts}
\usepackage{amssymb}
\usepackage{graphicx}

\newcommand{\wb}{\omega_{\mathrm{b}}}
\newcommand{\wi}{\omega_{\mathrm{imp}}}
\newcommand{\middlefig}{.25\textwidth}
\newcommand{\singlefig}{.5\textwidth}
\newcommand{\m}{m^*}

\begin{document}

\title{Breather Statics and Dynamics in Klein--Gordon
Chains with a Bend}

\author{J. Cuevas} \email{jcuevas@us.es}
\affiliation{Grupo de F\'{\i}sica No Lineal. Departamento de
F\'{\i}sica Aplicada I. ETSI Inform\'{a}tica. Universidad de
Sevilla. Avda. Reina Mercedes, s/n. 41012-Sevilla. Spain}

\author{P.G. Kevrekidis} \email{kevrekid@math.umass.edu}
\affiliation{Department of Mathematics and Statistics, University
of Massachusetts, Amherst, MA 01003-4515, USA.}

\date{\today}

\begin{abstract}
In this communication, we examine a nonlinear model with an
impurity emulating a bend. We justify the geometric interpretation
of the model and connect it with earlier work on models including
geometric effects. We focus on both the bifurcation and stability
analysis of the modes that emerge as a function of the strength of
the bend angle, but we also examine dynamical effects including
the scattering of mobile localized modes (discrete breathers) off
of such a geometric structure. The potential outcomes of such
numerical experiments (including transmission, trapping within the
bend as well as reflection) are highlighted and qualitatively
explained. Such models are of interest both theoretically in
understanding the interplay of breathers with curvature, but also
practically in simple models of photonic crystals or of bent
chains of DNA.
\end{abstract}



\maketitle

\section{Introduction}

In the last decade, intrinsic localized modes (ILMs), or discrete breathers
(DBs) as they are also termed, have become a topic of intense theoretical
and experimental investigation; see e.g., \cite{review} for a number of
recent reviews on the topic. Per their inherent ability to bottleneck
and potentially transport the energy in a coherent fashion,
such exponentially localized in space and periodic in time entities have
have come to be of interest in a variety of contexts. These range
from nonlinear optics and arrays of waveguides \cite{mora} to Bose-Einstein
condensates inside optical lattice potentials \cite{tromb} and from
prototypical models of nonlinear springs \cite{pesa} to Josephson
junctions \cite{alex} and dynamical models of the
DNA double strand \cite{Peybi}.

One of the playgrounds that have most recently been added to this long list of
ILM applications consists of nonlinear photonic crystal waveguides
and circuits \cite{kivshar}. In connection to this context, an issue
that becomes very relevant (see e.g. the models developed in \cite{kivshar})
is the interplay of nonlinearity and geometry as, typically in photonic
crystal waveguide arrays, two-dimensional or quasi-one dimensional
settings with bends \cite{kivshar,pgk1} become relevant.

We should note here that the interplay of nonlinearity and geometry
has been increasingly appreciated in the ILM literature. From the
long range interactions on a fixed curved substrate \cite{gaid},
to lattice-substrate feedback models \cite{pgk2} and from lattice junctions
with different masses \cite{bena}, to semi-circular, polymer-like
chains \cite{tsir} and geometrically motivated, bent models of DNA
\cite{ACMG01,ACG01,CPAR02,CAGR02}, the geometry can significantly
affect the static (inducing e.g., multi-stability) and dynamic
(causing e.g., a variety of outcomes in the ILM interaction with curvature)
properties of the relevant lattice model.

\begin{figure}
\begin{center}
     \includegraphics[width=\singlefig]{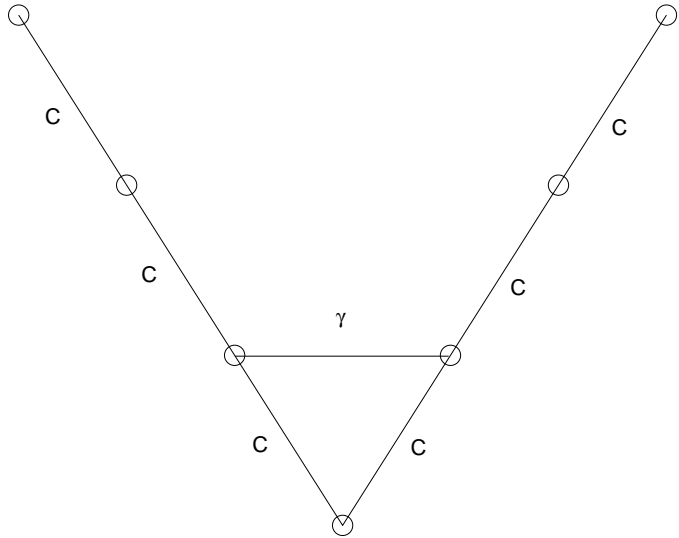}
\caption{Schematic presentation of a  bend in connection with the
discrete equation (\ref{eq:dyn}). The parameter $\gamma$ stands
for the bend-induced interaction between the next-nearest
neighbors (in the vicinity of the bend), whereas $C$ represents
the nearest neighbor interaction in the rectilinear
chain.}%
\label{fign1}
\end{center}
\end{figure}

In the present work, motivated by these studies we examine the
nonlinear Klein-Gordon variant of a model introduced recently
in the context of the discrete nonlinear Schr{\"o}dinger equation
(DNLS) in \cite{pgk1}. This is a prototypical model
emulating the geometry of a lattice bend, by the inclusion in
the vicinity of the bend of next-nearest neighbor interaction
due to the proximity of these neighbors in this context
(see e.g., Fig. \ref{fign1}).

We will examine this bend in the framework of a soft and a hard
Klein-Gordon interaction potential, namely the Morse and the
hard $\phi^4$ potentials respectively. Apart from the inherent
interest of the interaction, at the static as well as dynamical
level, of discrete breathers with the ``geometric impurity''
induced by the bend, the model may be relevant to a number
of applications. In particular, the DNLS serves as the envelope
wave equation (at the discrete level) for Maxwell's equation,
hence, for the photonic applications, it may be of
more interest to identify the properties of the corresponding
Klein-Gordon model. Furthermore, this may also serve as a simple
dynamical model for understanding the interaction of a denaturation
bubble in DNA  \cite{Peybi,RB} with the local helical geometry of the double
strand. This is similar in spirit to the earlier study of \cite{dauxois},
where the helicoidal geometry was argued to induce a non-nearest neighbor
interaction across the double strand. Notice, also, that we follow
a slightly different path than \cite{pgk1}, by examining not only
static but also dynamic properties of the model and, in particular,
the potential outcomes of the breather-bend interaction.

In the next section, we will present the model  equation of interest,
while in section 3, we will study its linear modes. In section 4,
we will examine the static properties of the two models,
by examining
the bifurcations of breathing modes as a function of the ``bend parameter''
$\gamma$. In section 5, the corresponding dynamics properties of
the breather-bend interaction will be presented. Finally, in section 6,
we summarize our findings and present our conclusions.

\section{The model}

In accordance with the above description, the Hamiltonian of the
Klein--Gordon chains of interest  will be given by:

\begin{equation}\label{eq:ham}
H=\sum_n\left(\frac{1}{2}\dot u_n+V(u_n)+\frac{1}{2} C\,
(u_{n+1}-u_n)^2 + \gamma(u_{n+1}-u_{n-1})^2\delta_{n,0}\right),
\end{equation}%
thus the bending point is located at the particle $n=0$ and
implies, as is graphically indicated in Fig. \ref{fign1}, a
geometrically induced coupling of next-nearest neighbors (nnn)
adjacent to the bend site.

We should clarify here the nature of the configuration and of the
dynamics. The displacements $u_n$ represent an ``external'' field
such as e.g., the
electric field in a array of optical waveguides or the stretching
of a base pair in a DNA chain \cite{Peybi}. The chain of Fig.
\ref{fign1} should be considered as being geometrically {\it
fixed} on the plane of the paper. The dynamics of the
displacements is transverse, i.e., perpendicular to the plane of
the bent chain. In such a setting the coefficient of the
interaction between adjacent sites is determined by the (fixed)
geometric proximity of the nodes constituting the chain. For a
derivation of the discrete model in the first setting and an
explanation of the dependence of the interaction coefficient on
the geometry of the configuration and the inter-nodal distance,
see e.g., \cite{christo}. In this context, it is natural to assume
that the effect of the bend (due to the geometric proximity of
sites $n=1$ and $n=-1$) will be to induce the nnn coupling between
them (cf. \cite{christo}).

$V(u)$ is
the on--site potential which is chosen to be of the (soft) Morse type,
$V(u)=(\exp(-u)-1)^2/2$ or of the (hard) $\phi^4$ type,
$V(u)=x^2/2+x^4/4$; $C$ is a coupling constant between nearest
neighbors (nn) whose interaction will be harmonic (i.e., connected by linear
springs) for the purposes of the present work.

From the Hamiltonian (\ref{eq:ham}), the following dynamical
equations can be obtained:

\begin{equation}\label{eq:dyn}
\ddot u_n+V'(u_n)+C \left( u_{n+1} + u_{n-1} -2 u_n \right) +
\gamma[(u_n-u_{n-2})\delta_{n,1}+(u_n-u_{n+2})\delta_{n,-1}]=0.
\end{equation}

A natural bifurcation/continuation parameter that we will use for the
purposes of our study is $\alpha\equiv\gamma/C$. The relevant
ratio is a natural measure of the relative strength of the
different neighbor interactions (the bend-based nnn one and the uniformly
distributed nn one). This ratio can also be interpreted geometrically
(in the sprit of Fig. \ref{fign1}). In particular, it can be
related to the wedge angle $\phi$ through the relation:
$\alpha=1/(2(1-\cos(\phi)))$. For the
nnn approximation to be realistic, $\phi$ must be
larger than $60^\mathrm{o}$, hence, equivalently, $\alpha<1$.
Notice that while the geometric interpretation of the moment prompts
us to typically examine the regime of $\alpha \in (0,1]$, it is
of inherent mathematical interest to examine broader parameter
ranges, and therefore some of our results below will be presented for
values of $\alpha$ outside this range.

\section{Linear modes}

Some of the properties of discrete breathers are related to the
existence or non-existence of linear localized modes. These modes
appear as the bend is introduced by modifying the coupling, which
is effectively equivalent to the introduction of an inhomogeneity
in the curvature at the bottom of the inter-site potential
\cite{CPAR02b}.

The linear modes can be obtained from the linearized form
(around the uniform state $u_n=0$) of the dynamical equations (\ref{eq:dyn}):

\begin{equation}\label{eq:linear}
\ddot u_n+\omega_o^2u_n+C(2u_n-u_{n-1}-u_{n+1})+
\gamma[(u_n-u_{n-2})\delta_{n,1}+(u_n-u_{n+2})\delta_{n,-1}]=0,
\end{equation}%
with $\omega_o^2=1$. The linear modes can be calculated using the
lattice Green's function \cite{BK91,MMW}. The
frequency of the localized linear modes (also referred to as
\emph{impurity modes}) is thus given by:

\begin{equation}\label{eq:lim}
\wi^2=\omega_o^2+\frac{C(2\alpha+1)^2}{2\alpha}=
\omega_o^2+\frac{(2\gamma+C)^2}{2\gamma},
\end{equation}%
and the inhomogeneity parameter is related to the frequency of the
impurity modes through the relation:

\begin{equation}\label{eq:lim2}
\alpha=\frac{\wi^2-\omega_o^2-2C\pm\sqrt{(\wi^2-\omega_o^2)(\wi^2-\omega_o^2-4C)}}
{4C},
\end{equation}%
where the plus sign corresponds to $\wi^2>\omega_o^2+4C$ and the
minus sign to $\wi<\omega_o$.

\begin{figure}
\begin{center}
\begin{tabular}{cc}
    \multicolumn{2}{c}{\includegraphics[width=\middlefig]{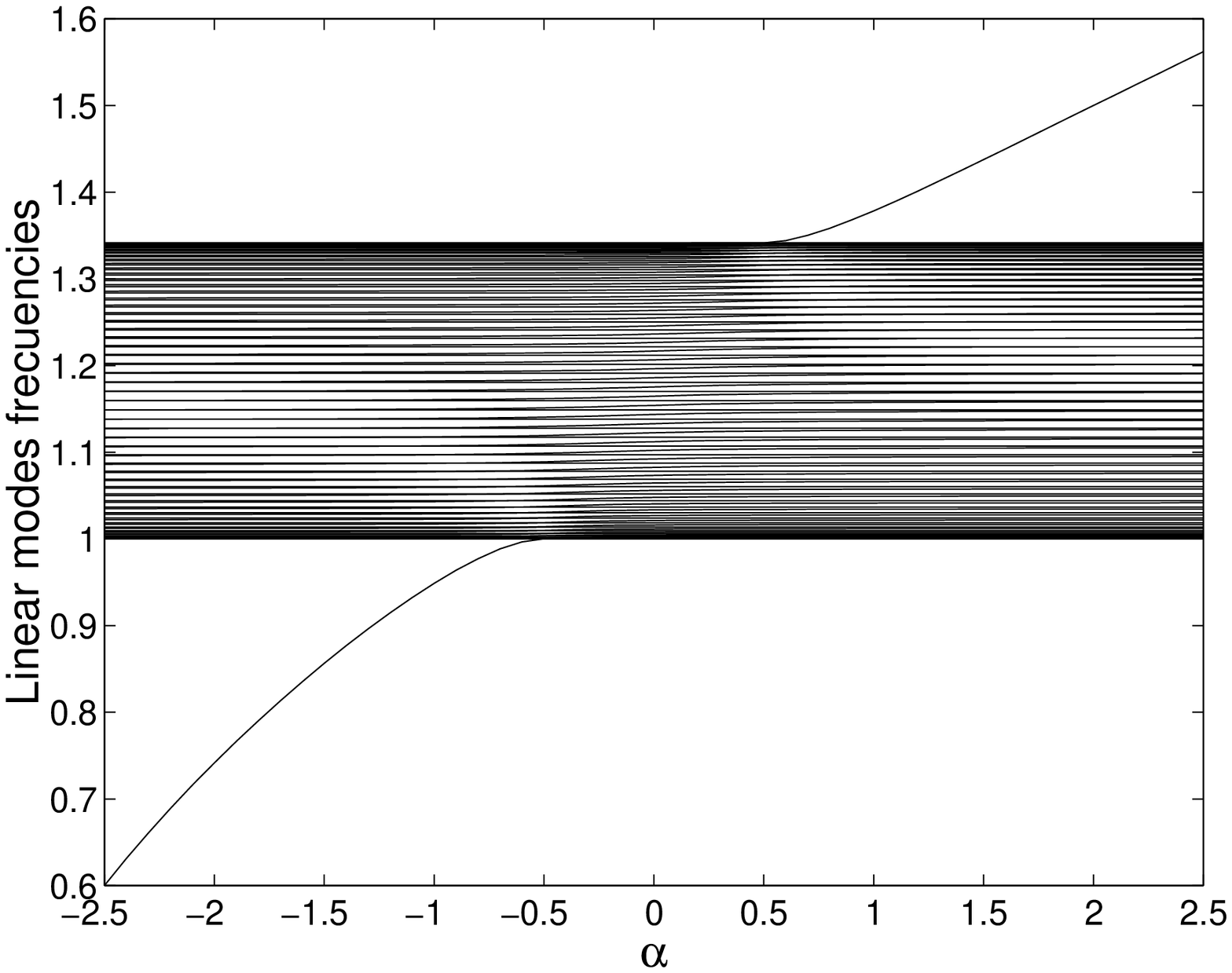}}\\
    \includegraphics[width=\middlefig]{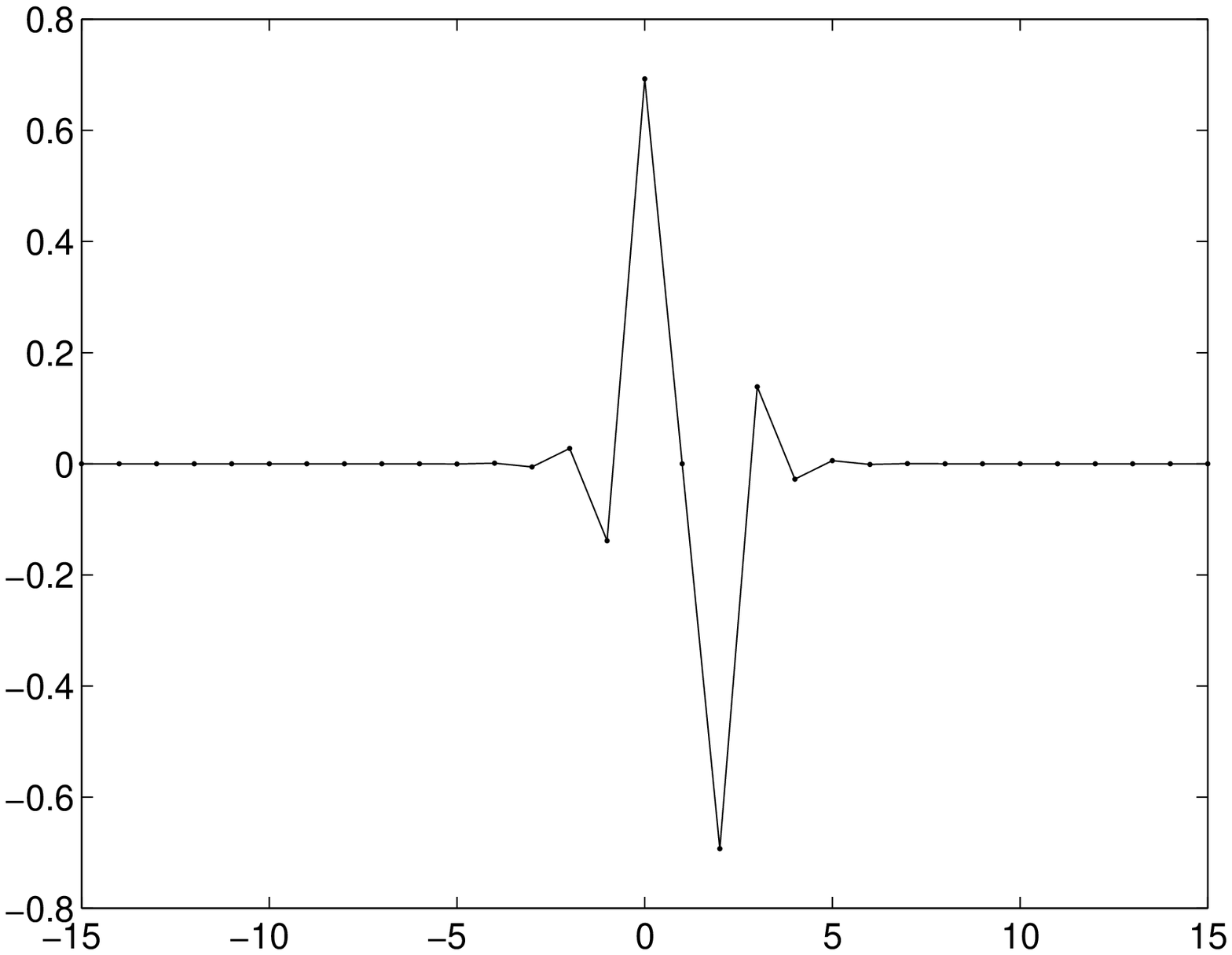} &
    \includegraphics[width=\middlefig]{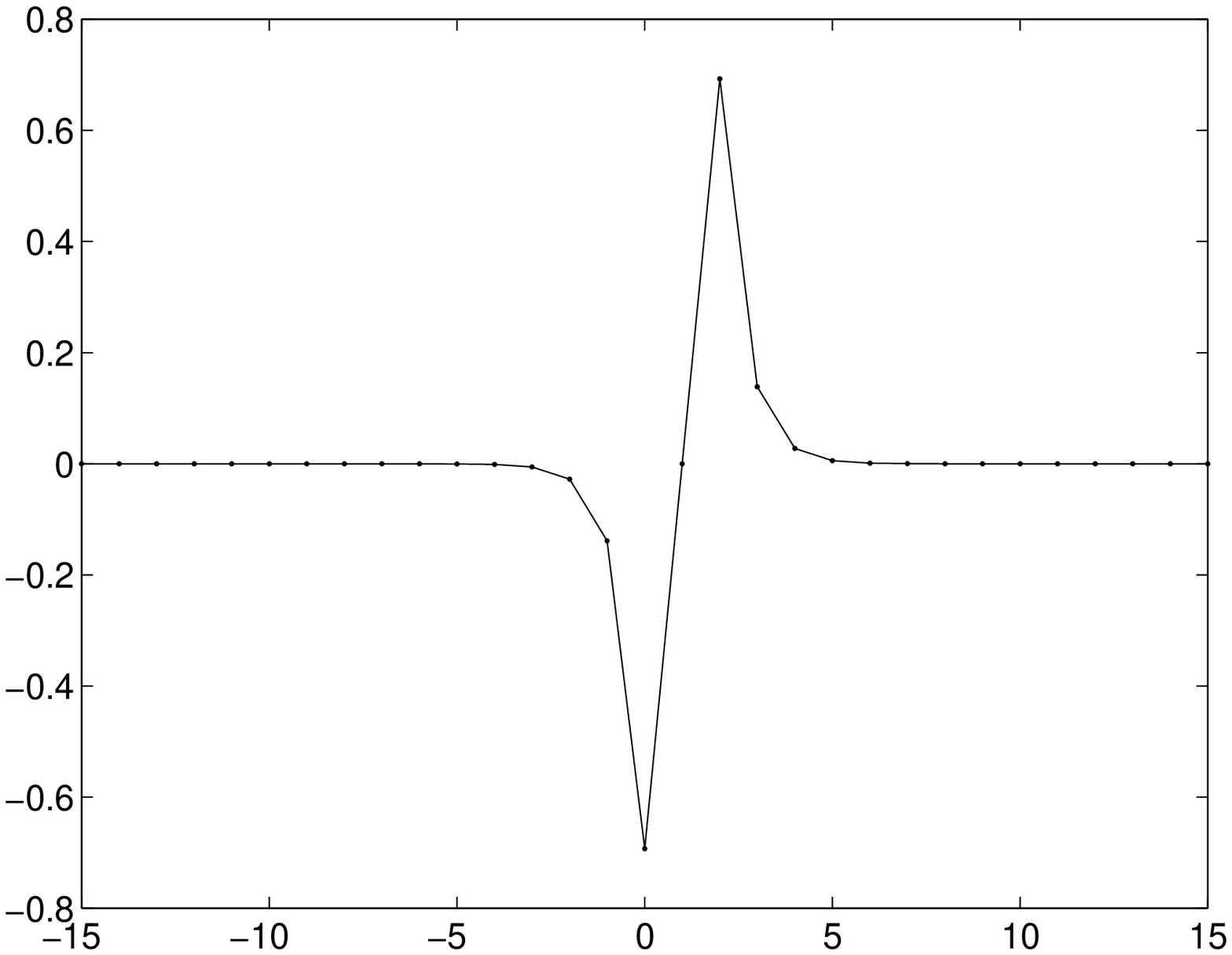}\\
\end{tabular}
\caption{(Top) Linear modes spectrum with respect to the bending
parameter $\alpha=\gamma/C$ for a coupling constant $C=0.20$.
(Bottom) Profile of the impurity mode for $\alpha>0$ (left) or
$\alpha<0$ (right) and $|\alpha|=2.5$. Note that the localization
of the impurity modes is non-negligible only for high values of $\alpha$.}%
\label{fig:linear}
\end{center}
\end{figure}

Figure \ref{fig:linear} shows the frequencies of the linear modes
as a function of $\alpha=\gamma/C$ and the profile of the impurity
modes. It can be observed that, for $\gamma>0$ the impurity mode
is above the phonon band and, in consequence, has a zigzag
vibrational pattern (i.e., a staggered mode).
If $\gamma<0$, the mode is below the band and
its sites oscillate in phase \cite{CPAR02b}. It is worth remarking that the
impurity modes are anti-symmetric modes whose central lattice site is
at rest.

\section{Static Results: Stationary breathers and their Bifurcations}

Discrete breathers can be calculated using well-known techniques
based on the concept of continuation from the
anti-continuous limit \cite{MA96}.
Furthermore, since the bending acts as an inhomogeneity, the system
has lost its integer shift invariance and the properties of the
breathers critically depend on the site where their center is located
(at $\gamma=0$). This
fact leads to the existence of saddle-node bifurcations through which
some of the solutions can disappear. Similar bifurcations have been
previously observed in bent chains of oscillators
\cite{ACMG01,ACG01}. This has also been the main focus, for DNLS type
settings, of \cite{pgk1}.

These bifurcations need a relatively high value of the coupling
(sufficient cross-talk between the neighboring sites) to be observed.
Otherwise, breather solutions will exist for very large intervals of
$\gamma$ in every site of the lattice. Hence we restrict ourselves to
$C$'s of the order of (typically) $0.1$ to render these bifurcations
tractable.

We now proceed to examine the bifurcation diagrams (as a function
of $\alpha$) for both soft and hard on-site potentials and a
harmonic inter-site potential.

\subsection{Morse potential}

\begin{figure}
\begin{center}
    \includegraphics[width=\singlefig]{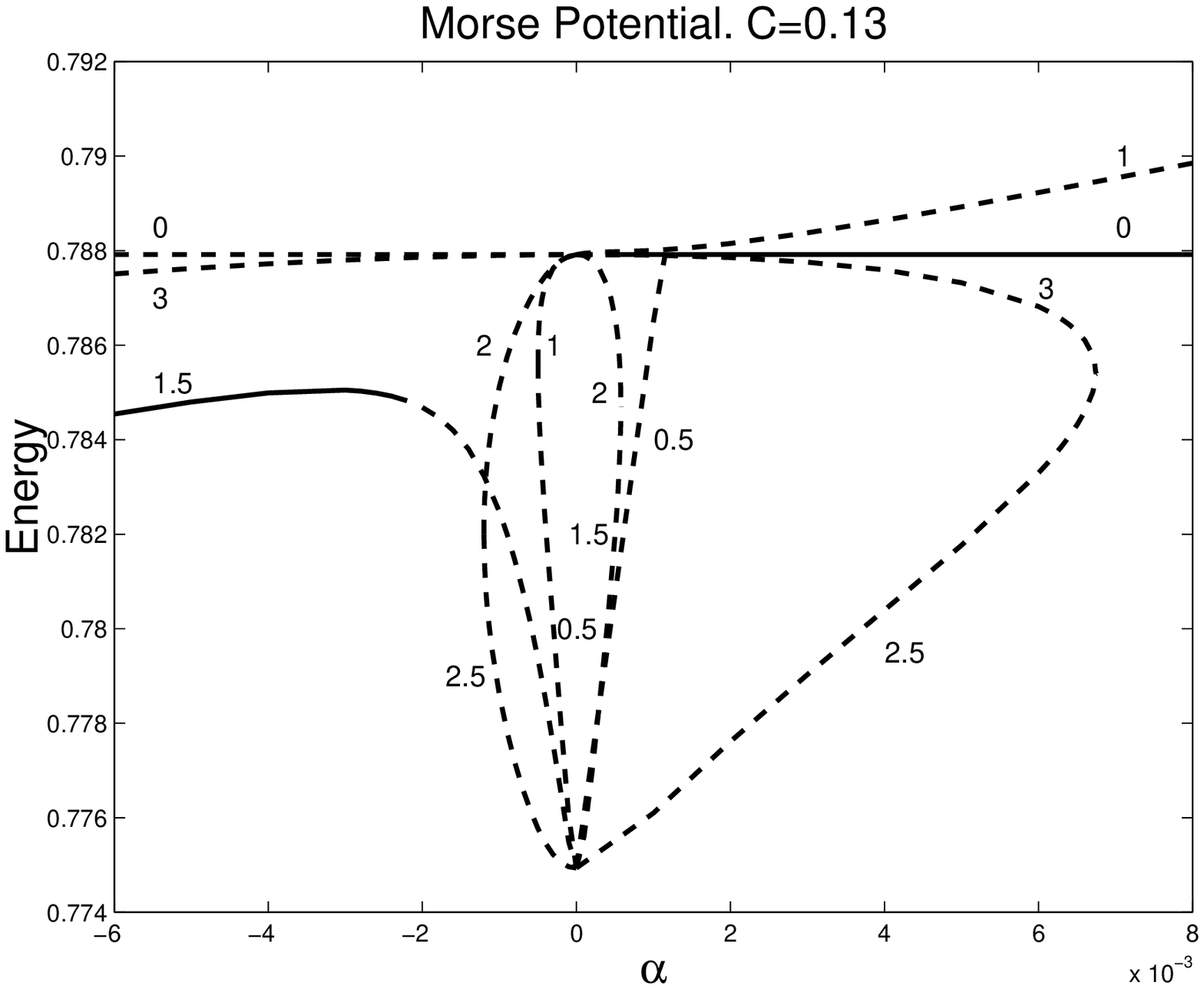}\\
    \includegraphics[width=\singlefig]{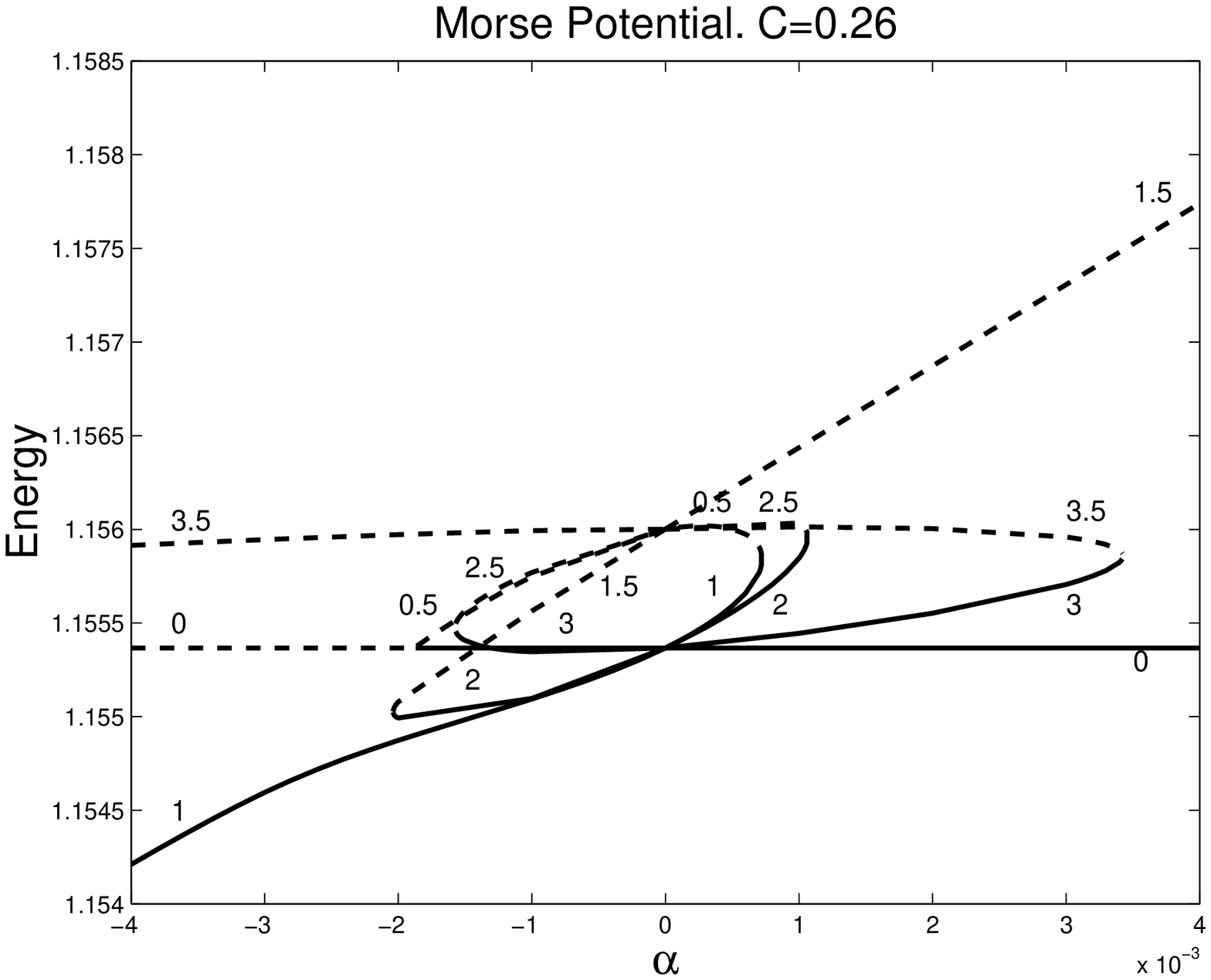}\\
\caption{Bifurcation diagram for the solutions with a Morse
on--site potential. The energy of the breathers is plotted as a
function of the bending parameter $\alpha=\gamma/C$. The coupling
constant is $C=0.13$ (top) or $C=0.26$ (bottom). The numbers
indicate the site where the solutions are centered; an integer
number corresponds to a site-centered solution, whereas a
half-integer number corresponds to a bond-centered solution.
Stable solutions (for an infinite system) are represented by full
lines whereas unstable solutions are
represented by dashed lines.}%
\label{fig:bifMorse}
\end{center}
\end{figure}

\begin{figure}
\begin{center}
\begin{tabular}{cc}
    (a) & (b) \\
    \includegraphics[width=\middlefig]{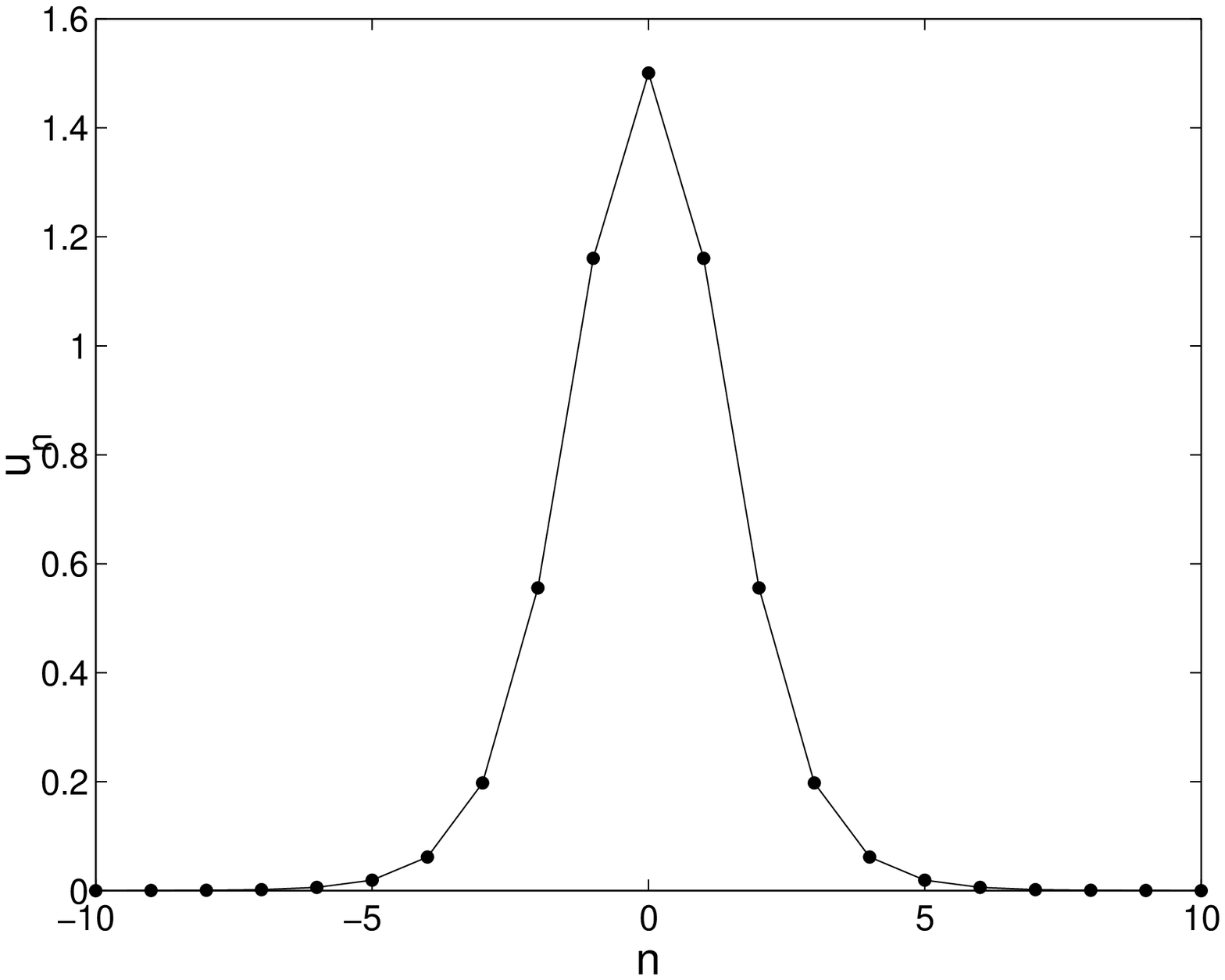} &
    \includegraphics[width=\middlefig]{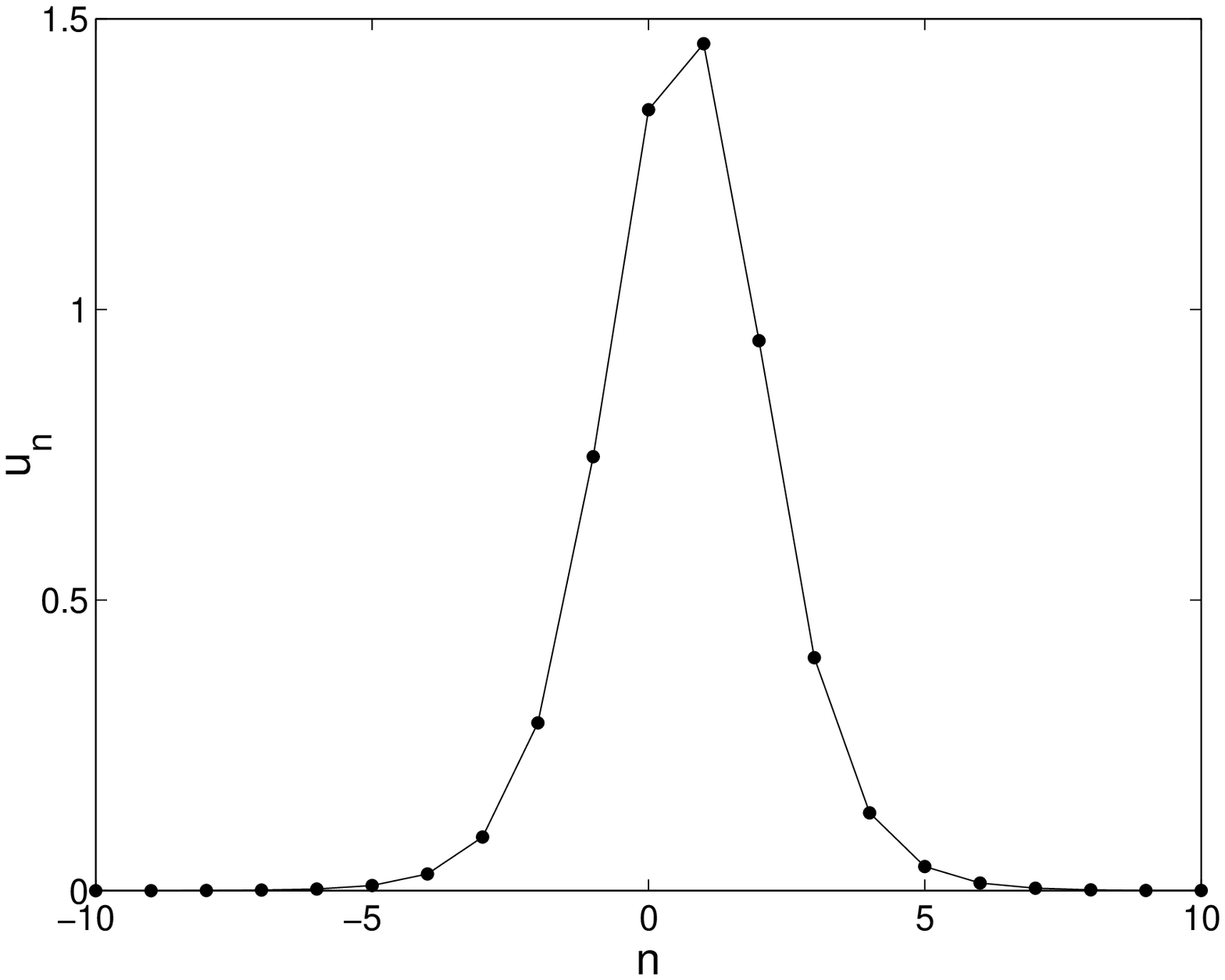}\\
    (c) & (d) \\
    \includegraphics[width=\middlefig]{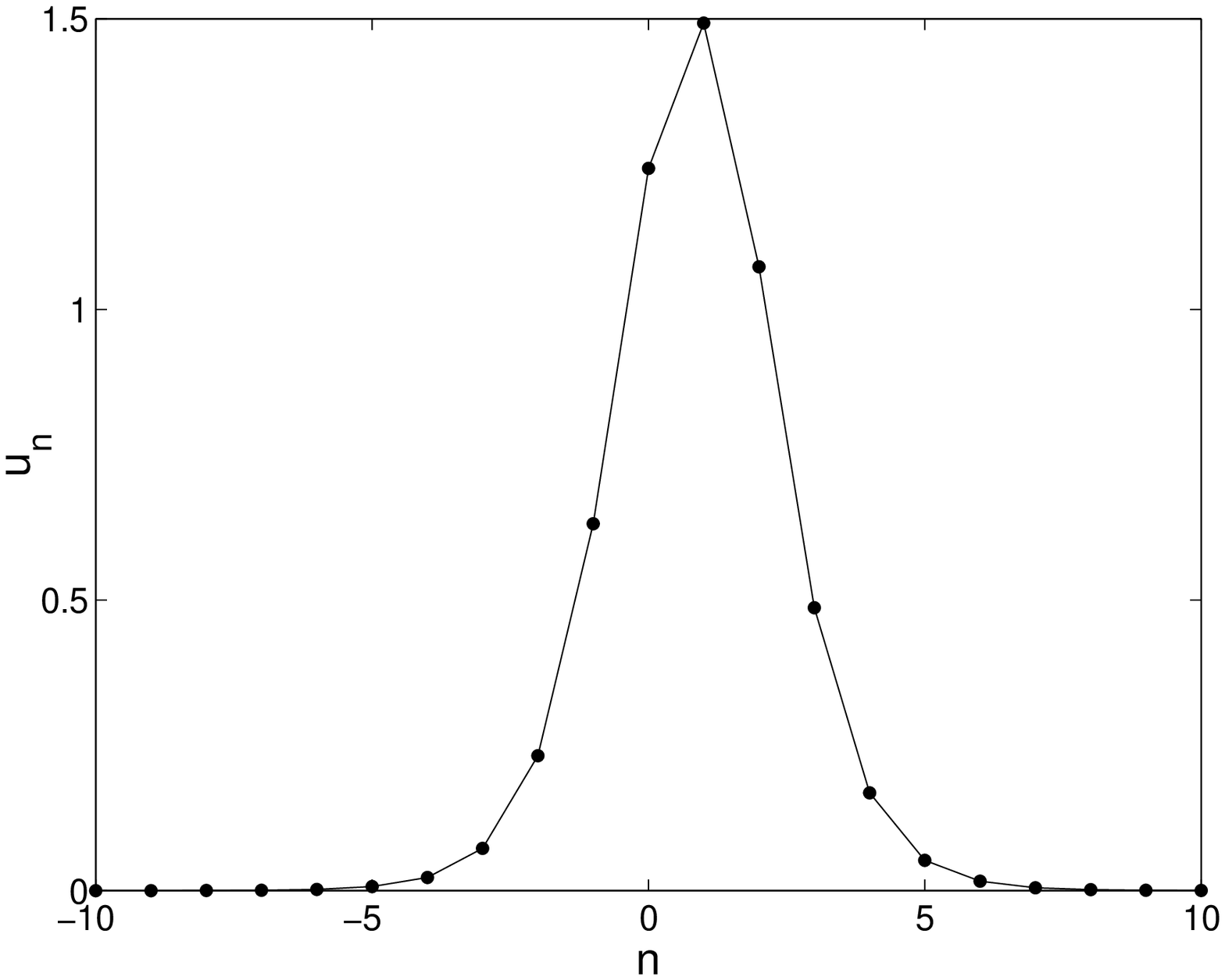} &
    \includegraphics[width=\middlefig]{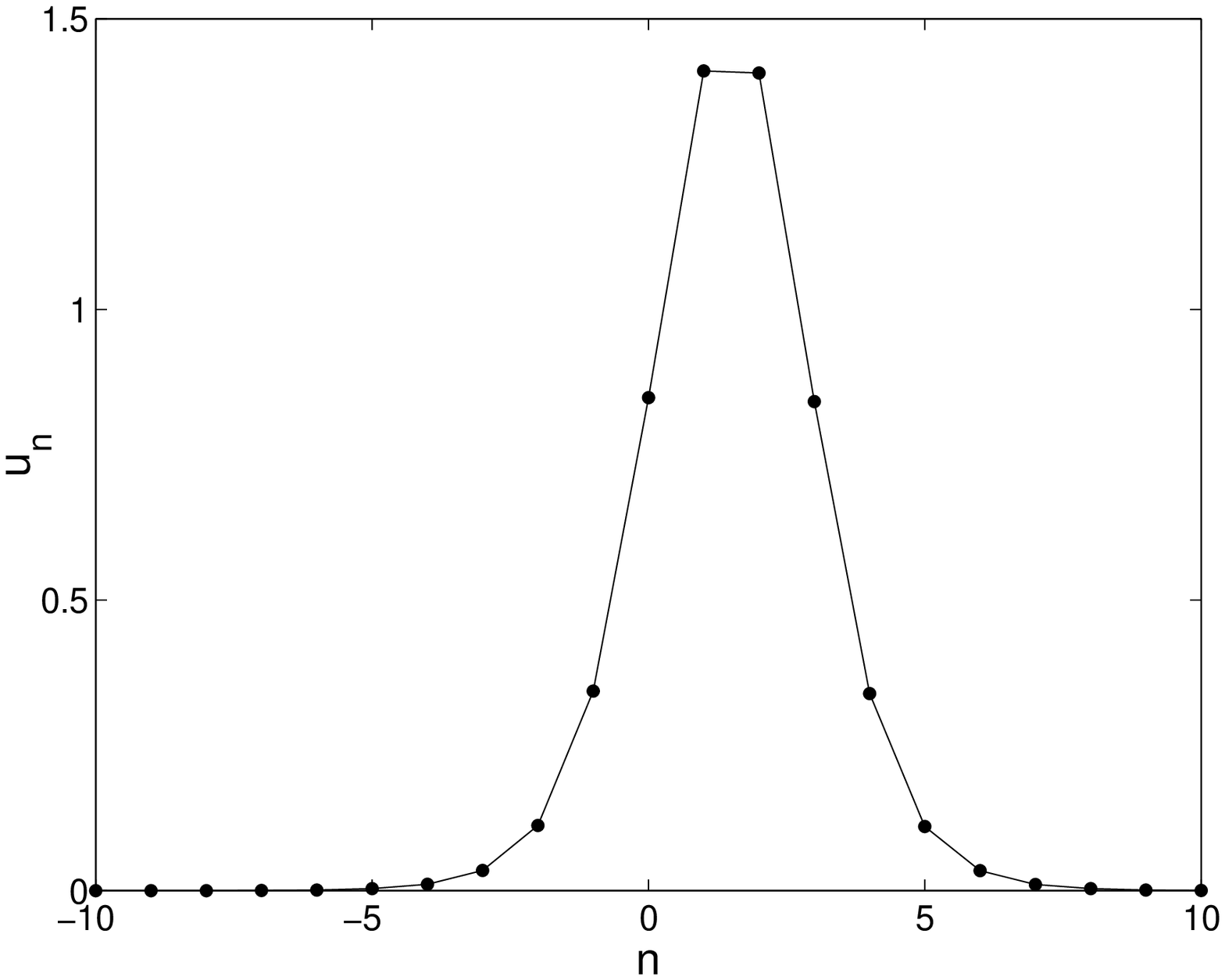}\\
\end{tabular}
\caption{Spatial profiles of the breathers centered at $n=0$ (a),
$n=0.5$ (b), $n=1$ (c) and $n=1.5$ (d) for $\alpha=0.0005$. The
results correspond to a chain of oscillators with an on--site
Morse potential with parameters $\wb=0.8$ and $C=0.26$.}%
\label{fig:brMorse}
\end{center}
\end{figure}

Figure \ref{fig:bifMorse} shows the bifurcation diagrams for the
case of a Morse on--site potential and couplings $C=0.13$ and
$C=0.26$ and frequency $\wb=0.8$, and Figure \ref{fig:brMorse}
shows the spatial profile of the solutions corresponding to
a number of branches for $\alpha=0.0005$.

For $C=0.13$, the branch corresponding to the breathers centered
at $n=0$
merges with the breathers originally centered at $n=0.5$ at
$\alpha\approx0.00115$. The branches centered at $n=1.5$ and $n=2$
disappear through a saddle-node bifurcation at $\alpha\approx0.00057$. This
phenomenon is also observed for the branches centered at $n=2.5$
and $n=3$ at $\alpha\approx0.00674$. For $\alpha<0$ the
annihilations are observed between different branches of
solutions. In particular, the branches centered at $n=0.5$ and
$n=1$ annihilate at $\alpha\approx-0.00049$ and the ones centered
at $n=2$ and $n=2.5$ cease to exist through the saddle-node
bifurcation occurring at $\alpha\approx-0.00120$. It should,
however, be noted that some branches (see e.g., the branch
centered at $n=1$ for $\alpha>0$ or the one centered $n=-1.5$ for
$\alpha<0$) never annihilate. It is also interesting to note that
all of these branches of solutions are unstable, as both
site-centered and bond-centered breathers are unstable in the
straight chain (for this value of the nn coupling). However, the
branch resulting when the centered at $n=0$ and $n=0.5$ merge is
stable.

The branch centered at $n=1.5$ becomes stable for
$\alpha\lesssim0.002$. This change of stability has its origin in
a pair of Floquet multipliers that collide in the unit circle at
$\lambda=1$. Notice that this behavior, unlike the typical
branches discussed above does not involve a saddle-node
bifurcation (or for that matter an exchange of stability with
another static breather branch).

For $C=0.26$ the saddle-node bifurcations occur between different
solutions from the case considered above. For $\alpha>0$, the
branches centered at $n=0.5$ and $n=1$ collide and disappear at
$\alpha\approx0.00072$; the branches centered at $n=2$ and $n=2.5$
cease to exist at $\alpha\approx0.00106$ while the ones centered at
$n=3$ and $n=3.5$ terminate at $\alpha\approx0.00342$. The
branch centered at $n=0$ merges with the branch centered at
$n=0.5$ for $\alpha<0$ and in particular for
$\alpha\approx-0.00187$.

In addition, the branches centered at $n=1.5$ and $n=2$ annihilate
at $\alpha\approx-0.00204$ and the ones centered at $n=2.5$ and
$n=3$ collide at $\alpha\approx-0.00157$. It is interesting to
observe that similarly to what was found in \cite{pgk1}, in a
different setting, there is an asymmetric mode that persists for
$\alpha < 0$, whose energy is lower than the symmetric bend mode.
Hence, for $\alpha<0$ in this case, we observe a symmetry breaking
effect that leads to an asymmetric ground state of the system.
Notice that similar asymmetric modes can be found (and identified
to be potentially stable) in continuum models with localized
impurities; see e.g., \cite{andrey}.

\begin{figure}
\begin{center}
\begin{tabular}{cc}
    (a) & (b) \\
    \includegraphics[width=\middlefig]{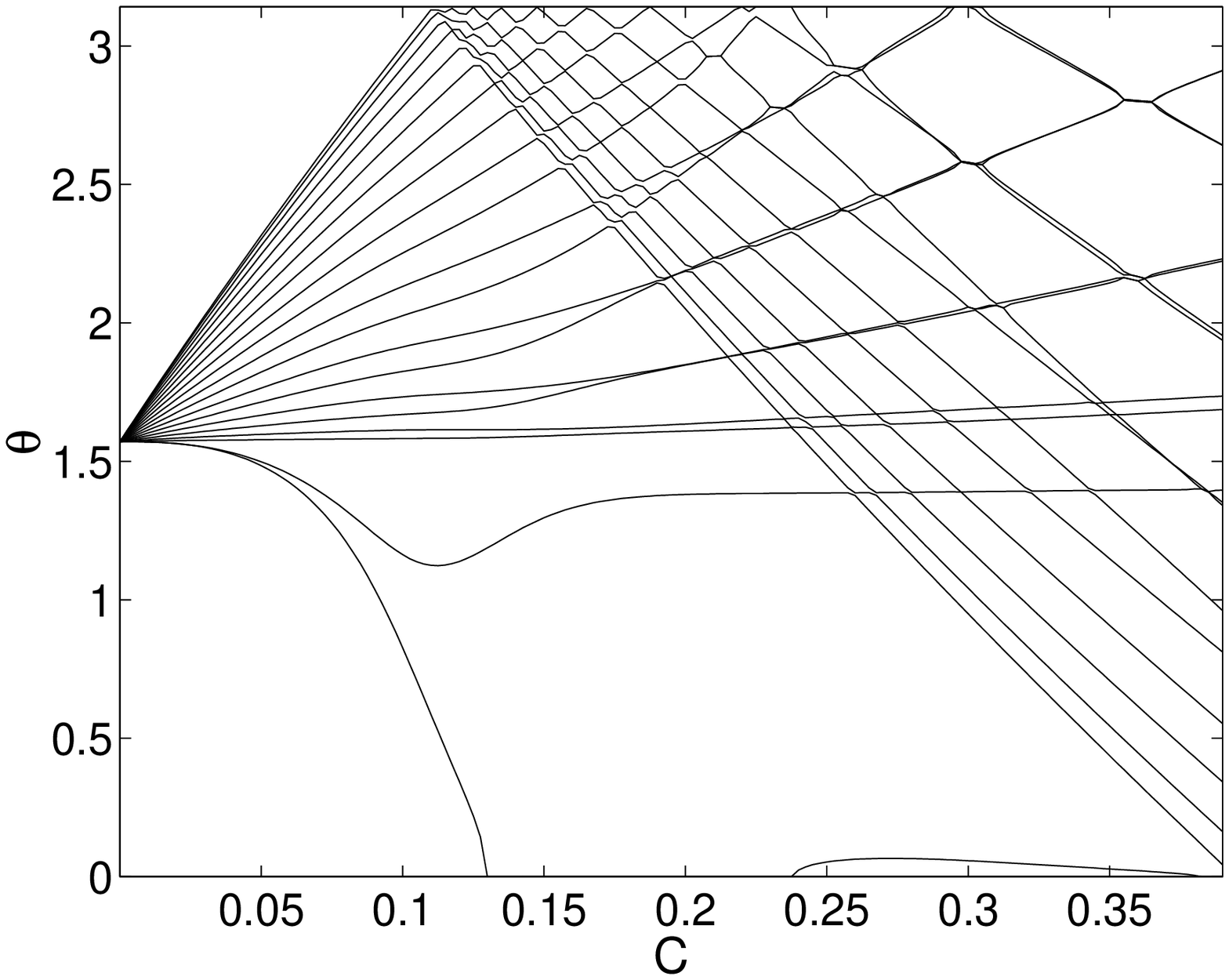} &
    \includegraphics[width=\middlefig]{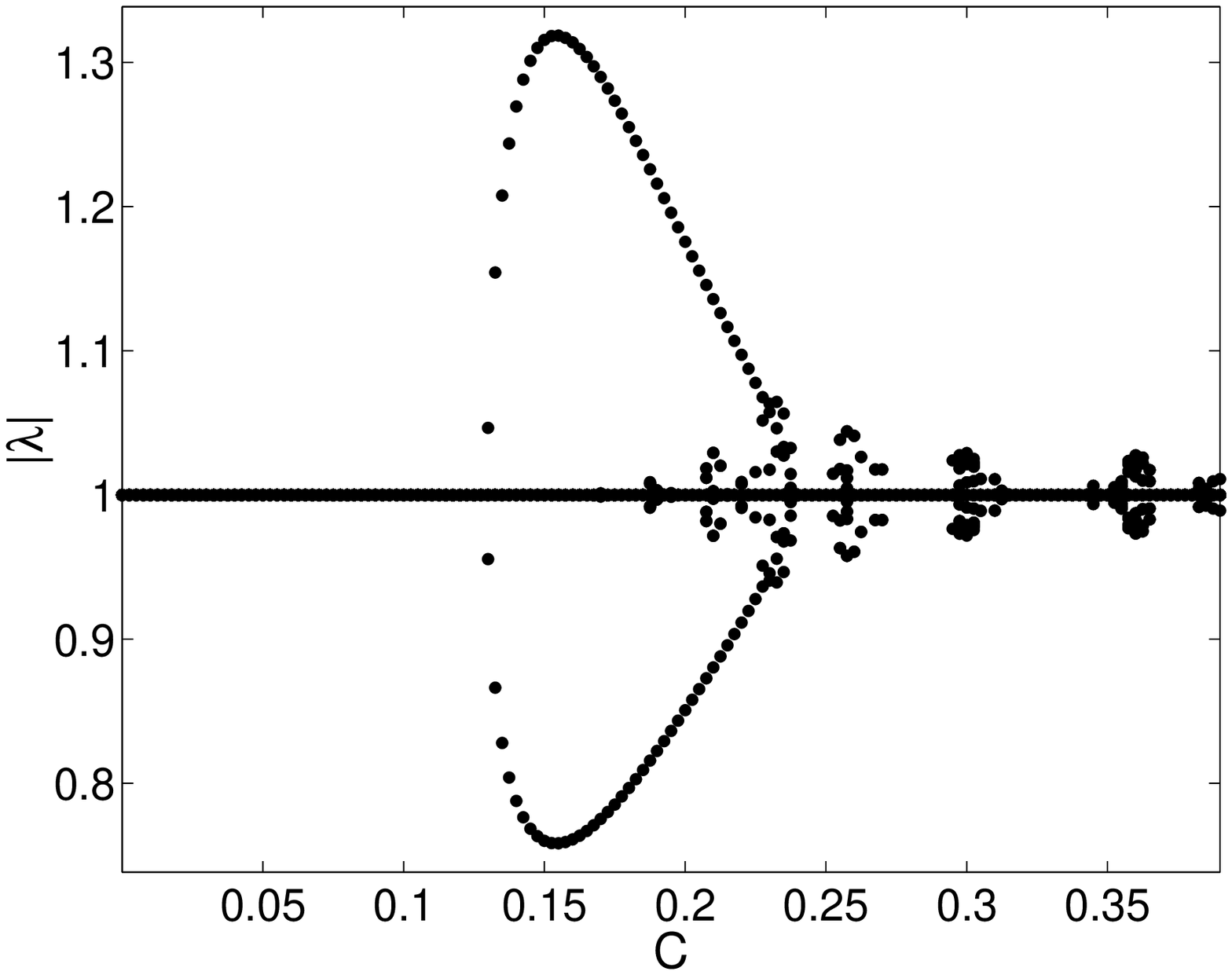}\\
    (c) & (d) \\
    \includegraphics[width=\middlefig]{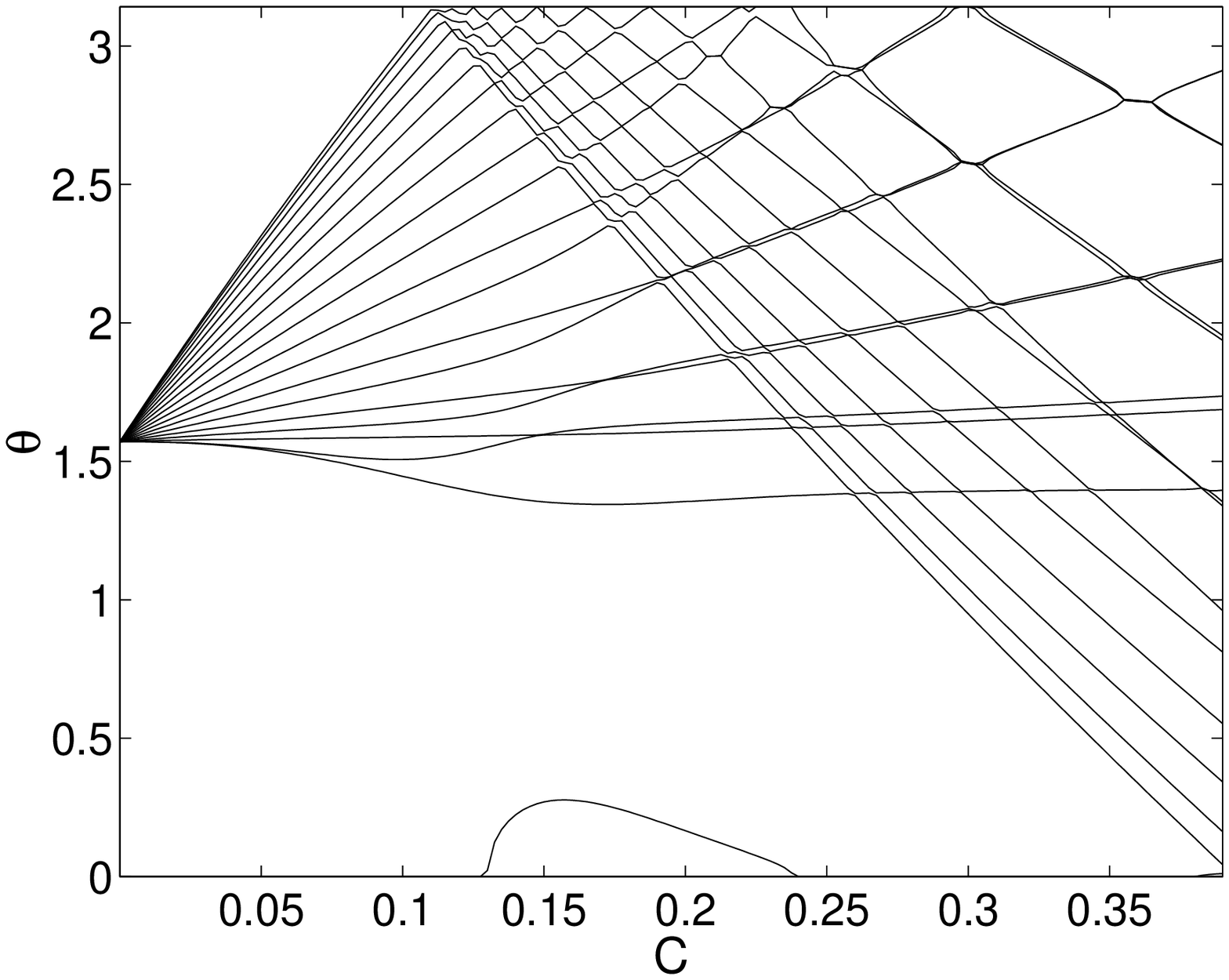} &
    \includegraphics[width=\middlefig]{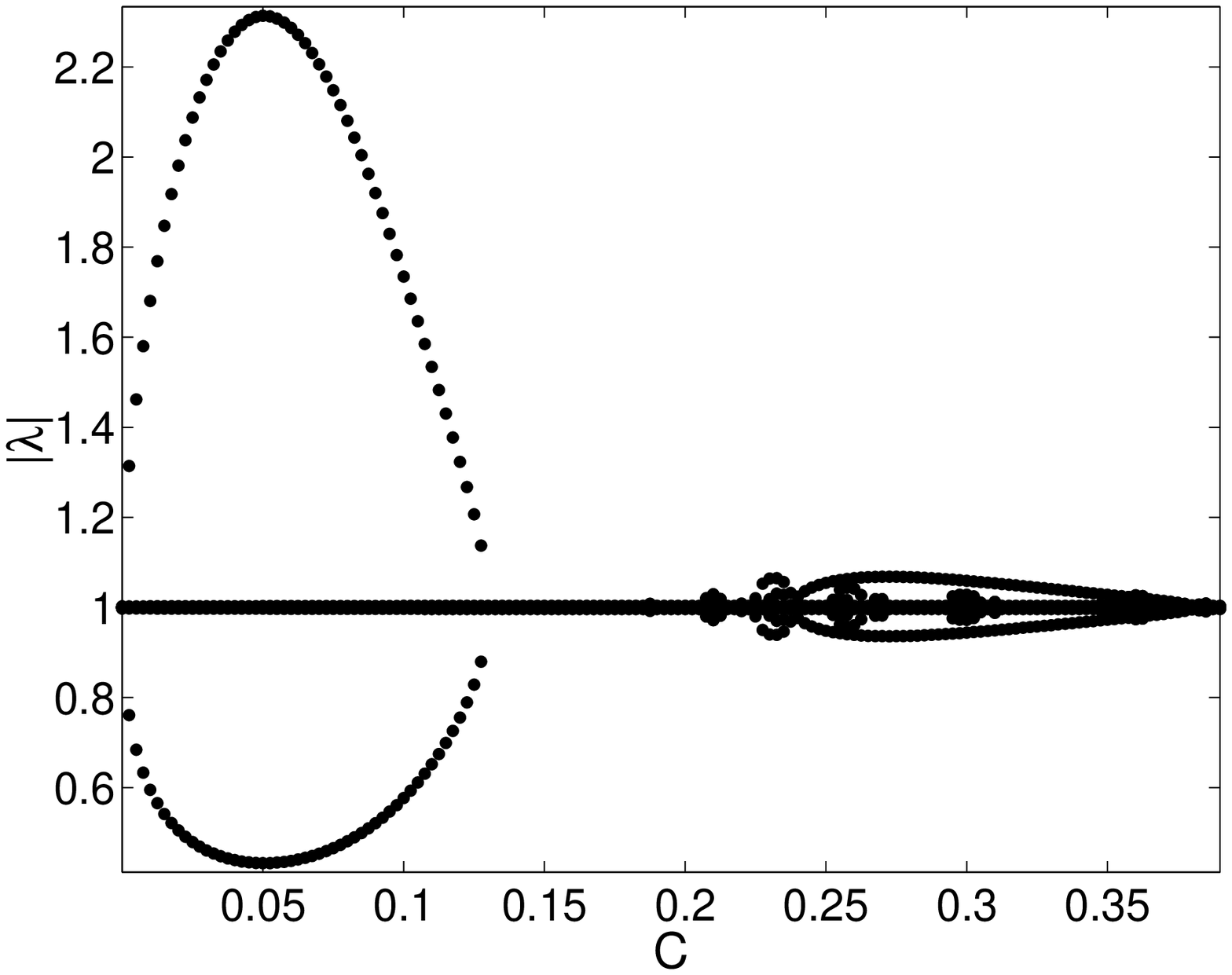}\\
\end{tabular}
\caption{Evolution with respect to the coupling constant of the
Floquet multiplier
arguments $\theta$ (left) and the modulus of the Floquet
multipliers (right) for a 1-site (top) and a 2-site (bottom)
breather with Morse on--site potential with frequency $\wb=0.8$.
The 1-site breather is stable for $C\in(0,0.1297)$ and for
$C\in(0.2382,0.4)$ it recovers its stability except for
size-dependent instability bubbles via oscillatory and subharmonic
bifurcations. The 2-site breather is stable for
$C\in(0.1300,0.2381)$.}%
\label{fig:floquet}
\end{center}
\end{figure}

It should be noted here that for this larger value of the
coupling, the branches corresponding to site centered solutions
are stable for an infinite lattice, whereas the bond-centered
solutions are unstable for the rectilinear chain. This stability
is inherited by the branches of the bent chain. Motivated by the change
of stability at the ``un-bent'' limit ($\gamma=0$), occurring as a
function of $C$ (e.g., notice above the different stability of
this limit for $C=0.13$ and $C=0.26$), we briefly study the
stability of the rectilinear chain limit, as a function of $C$. To
illustrate the stability of the 1-site and 2-site modes in the
rectilinear chain (discussed in part in \cite{CAGR02}), we examine
their Floquet multipliers, both by means of the argument angles as
well as giving their absolute values in Fig. \ref{fig:floquet}.
Notice, in particular, for the 1-site modes that there are two
inversions of stability, namely one at $C\approx0.1297$ and
another for $C\approx0.2382$. For $C>0.2382$, the 1-site modes are
stable but with interspersed size-dependent windows of instability
(see e.g., \cite{MA98,Cuevas,cretegny}).

\begin{figure}
\begin{center}
\begin{tabular}{cc}
    (a) & (b) \\
    \includegraphics[width=\middlefig]{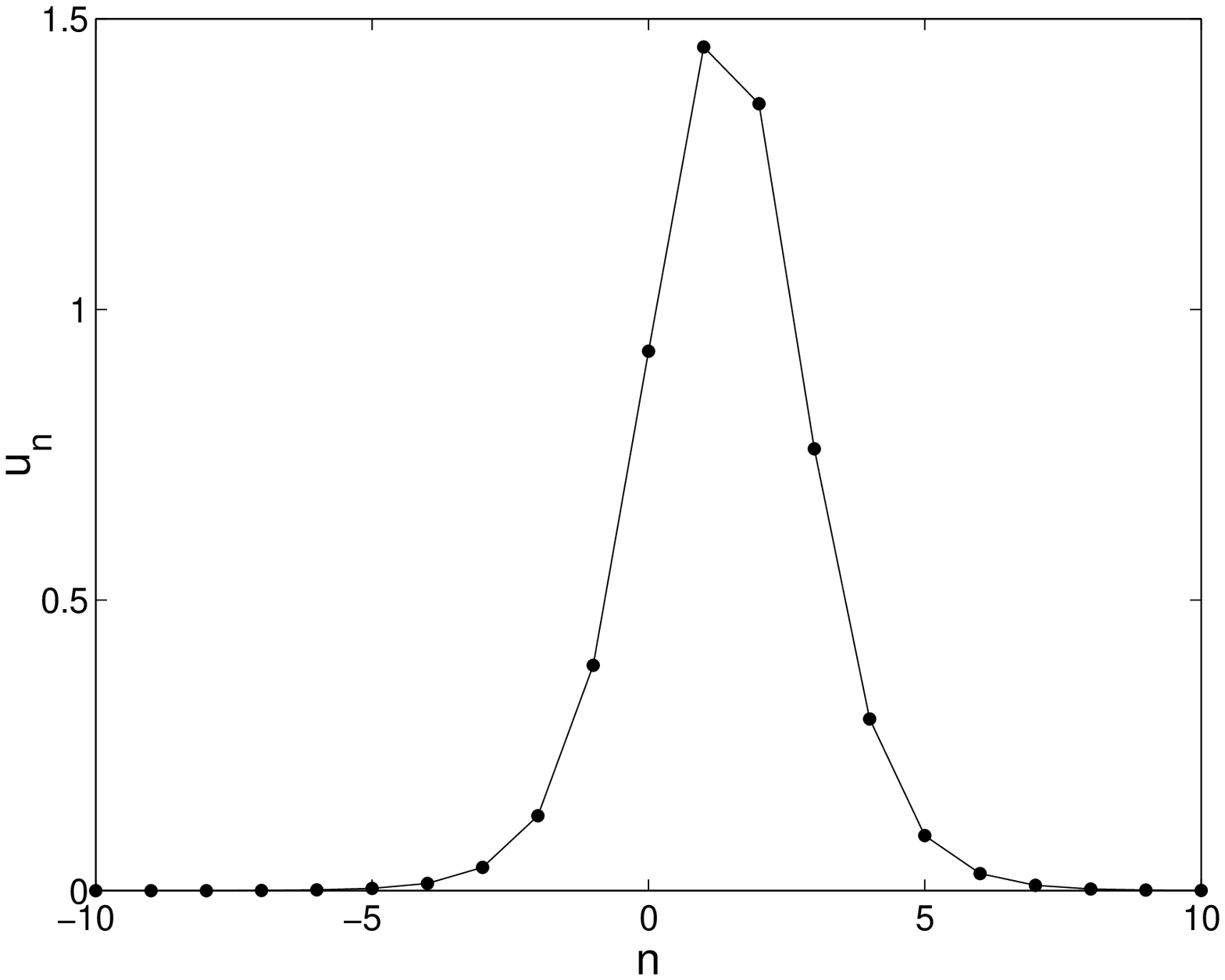} &
    \includegraphics[width=\middlefig]{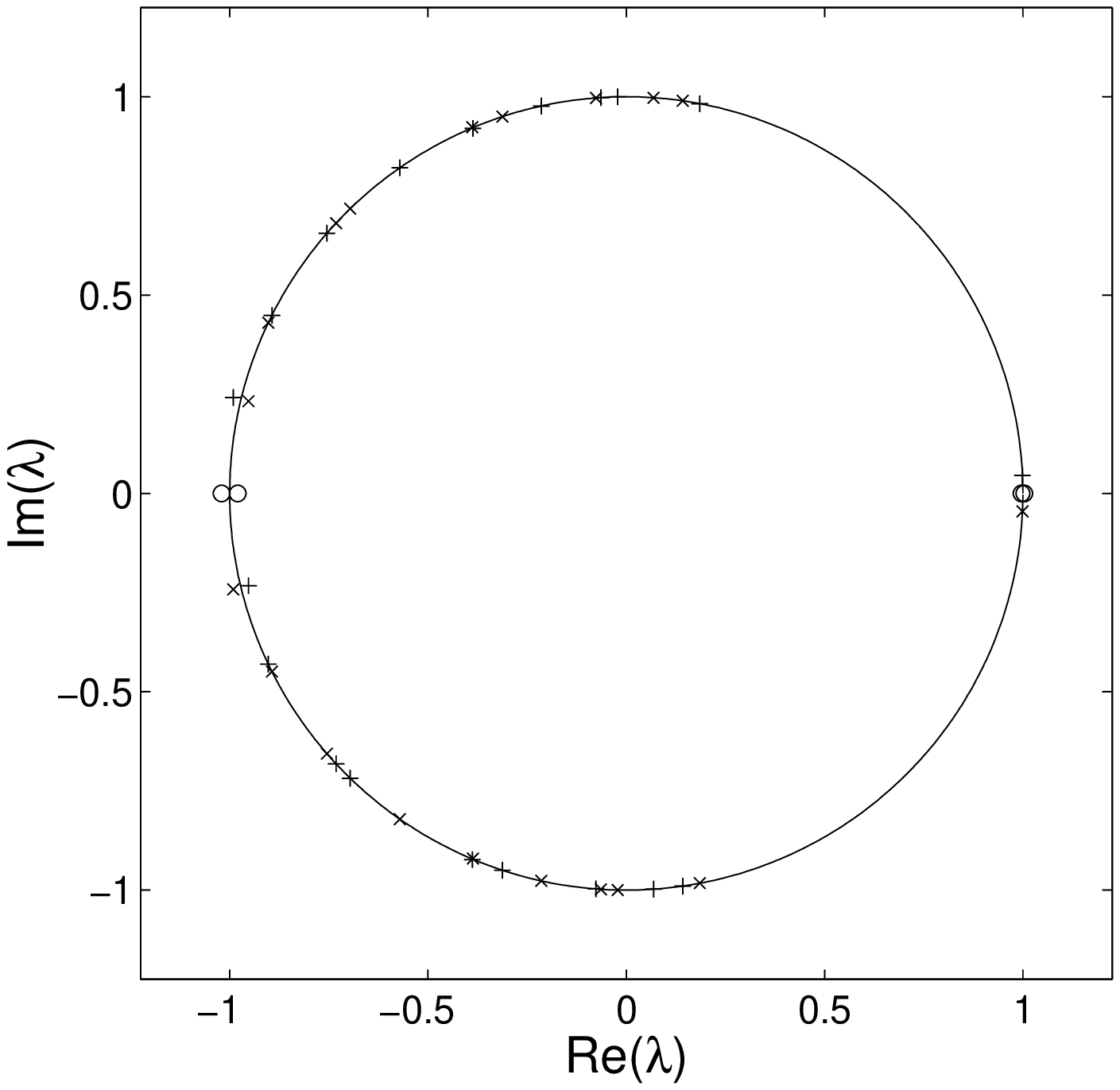}\\
\end{tabular}
\caption{(a) Spatial profiles of the breathers centered at $n=1$
for $\alpha=-0.003$ and the same characteristics to those of
Figure \ref{fig:brMorse}. (a) shows the spatial profile of
unstable solution;
(b) represents the Floquet multipliers corresponding to this state.
The instabilities that can be appreciated near
$\theta=\pi$ are due to the collision of extended eigenmodes and
disappear in the case of an infinite lattice.}%
\label{fig:stabMorse}
\end{center}
\end{figure}

Finally, let us mention that the branch that originates by the
merging of the $n=0$ and $n=0.5$ branches is unstable. Figure
\ref{fig:stabMorse} shows an example of the spatial profile and the
Floquet eigenvalues of an unstable state for $\alpha<0$, namely a breather
(originally) centered around $n=1$. Notice however that the instability
in this case is a result of the finite size of the lattice.

\subsection{$\phi^4$ potential}

The most significant difference of the $\phi^4$ case
with respect to the Morse one is
that in the former (i.e., for a hard potential), the breathers
bifurcate from the top of the continuous spectrum and as a result
the corresponding spatial profiles are staggered \cite{review}.
Hence, in this case the main solutions of interest are 1-site or
2-site breathers with the adjacent sites oscillating in
anti-phase.

\begin{figure}
\begin{center}
    \includegraphics[width=\singlefig]{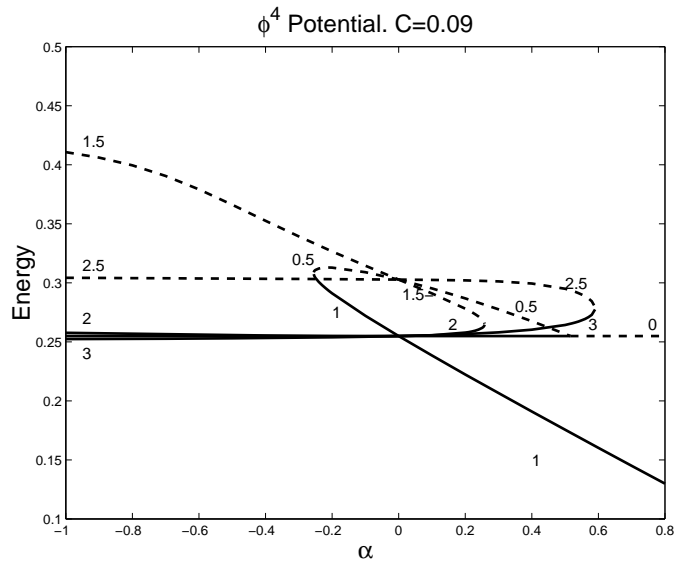}\\
\caption{Bifurcation diagram for the solutions with a $\phi^4$
on--site potential. The energy of the breathers is plotted as a
function the bending parameter $\alpha=\gamma/C$. The numbers
indicate the site where the solutions are centered; the
half-integer numbers actually correspond to 2-site breathers, but
the notation has been  kept in consonance with Figure
\ref{fig:bifMorse}. Stable solutions are represented by full lines
whereas unstable solutions are represented
by dashed lines.}%
\label{fig:bifphi4}
\end{center}
\end{figure}

The hard $\phi^4$ potential has been considered with a
coupling constant $C=0.09$ and solutions with frequency
$\wb=1.2$. For these parameter values (notice that $C$ is comparable
to the values used in Fig. \ref{fig:bifMorse}), the saddle-node bifurcations
of interest occur at considerably larger values of $\alpha$ for the
present model in comparison to the Morse case;
see e.g., Fig.
\ref{fig:bifphi4}. In particular, the branch of the on-site breathers
centered at $n=0$ merges with the inter-site branch
centered at $n=0.5$ at $\alpha\approx0.516$. Similarly, the on-site
solutions centered at $n=2$ and $n=3$ collide (and disappear)
 with the corresponding
inter-site ones centered at $n=1.5$ and $n=2.5$ at
$\alpha\approx0.258$ and $\alpha\approx0.590$, respectively.
However, in this case, the branch centered at $n=1$, becomes the
asymmetric, lowest energy, stable ground state that persists for
any value of $\alpha$. For $\alpha<0$, the on-site breathers
centered at $n=1$ annihilate with the inter-site ones centered at
$n=0.5$ at $\alpha\approx-0.255$. Notice that here also on-site
solutions are stable, while inter-site ones are unstable for
$\gamma=0$, but the branch emerging from the $n=0$ and $n=0.5$ is
unstable.

\begin{figure}
\begin{center}
\begin{tabular}{cc}
    (a) & (b) \\
    \includegraphics[width=\middlefig]{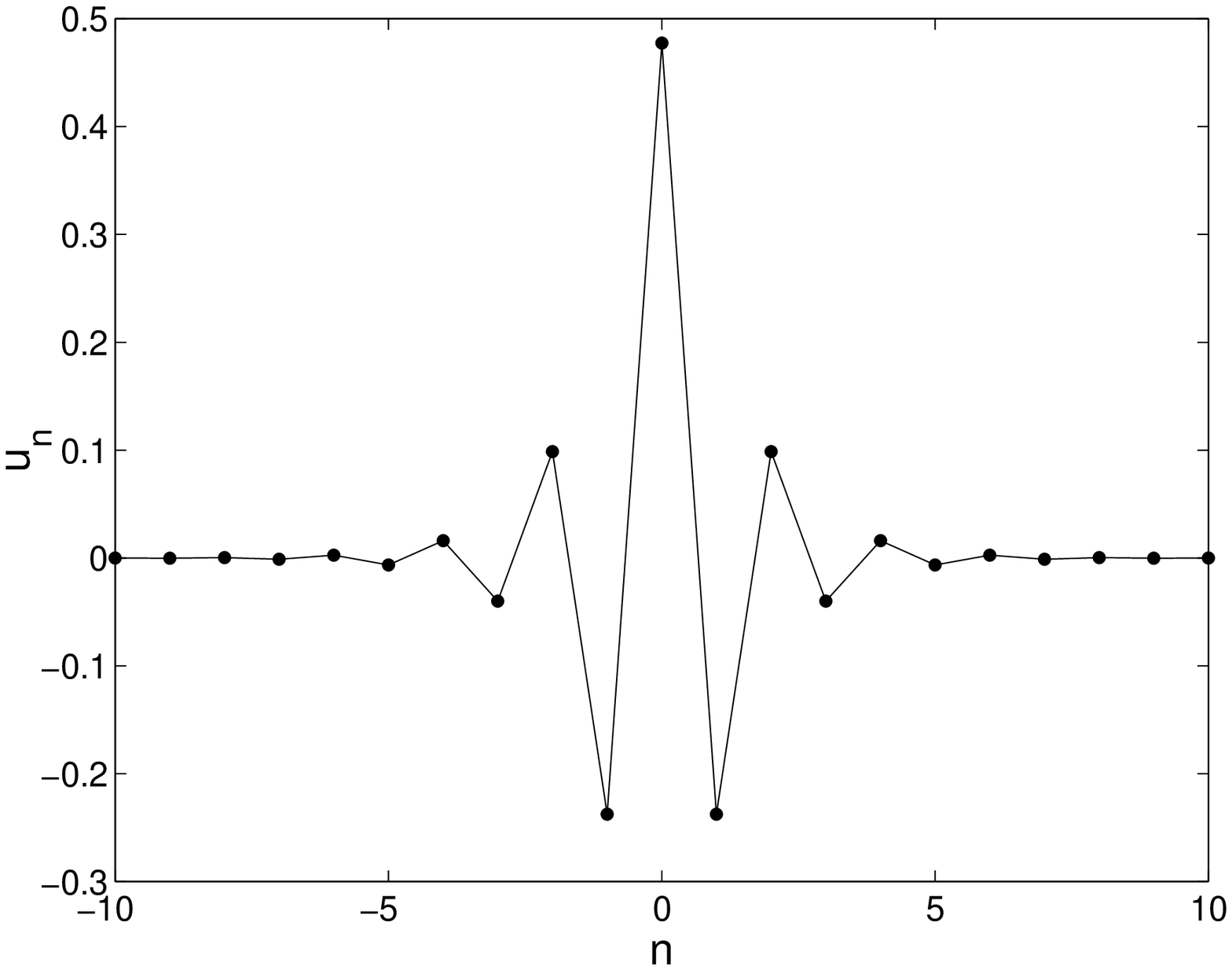} &
    \includegraphics[width=\middlefig]{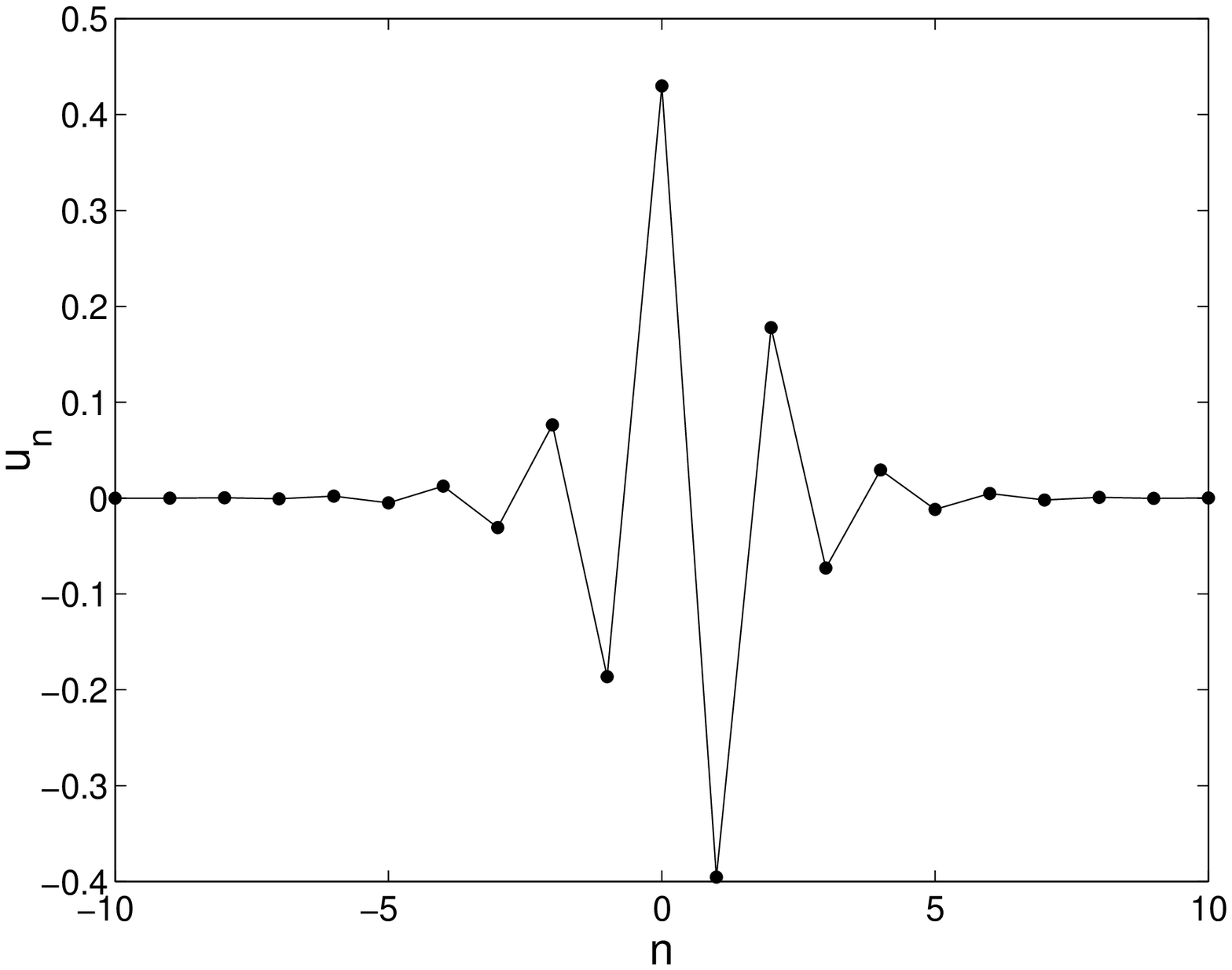}\\
    (c) & (d) \\
    \includegraphics[width=\middlefig]{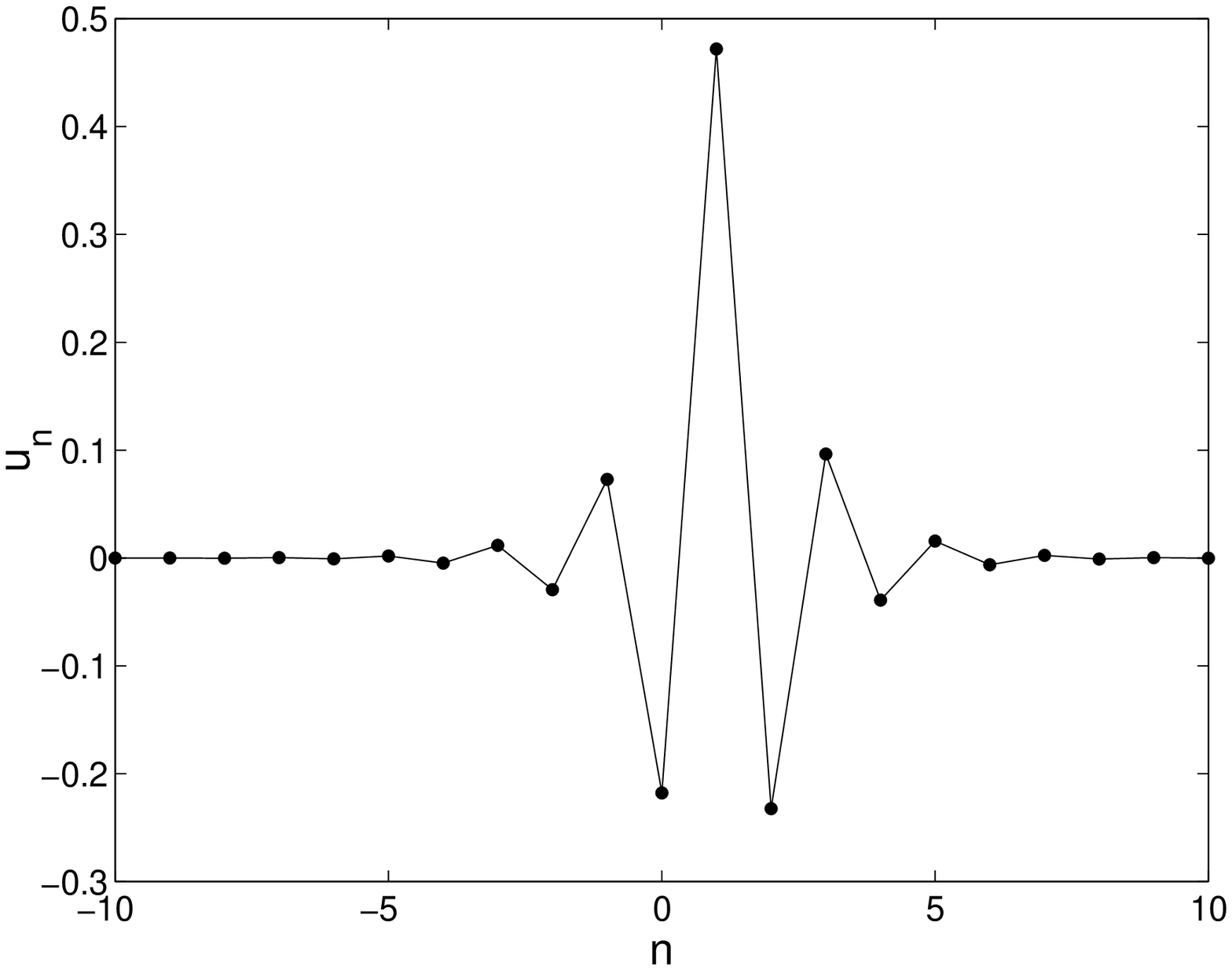} &
    \includegraphics[width=\middlefig]{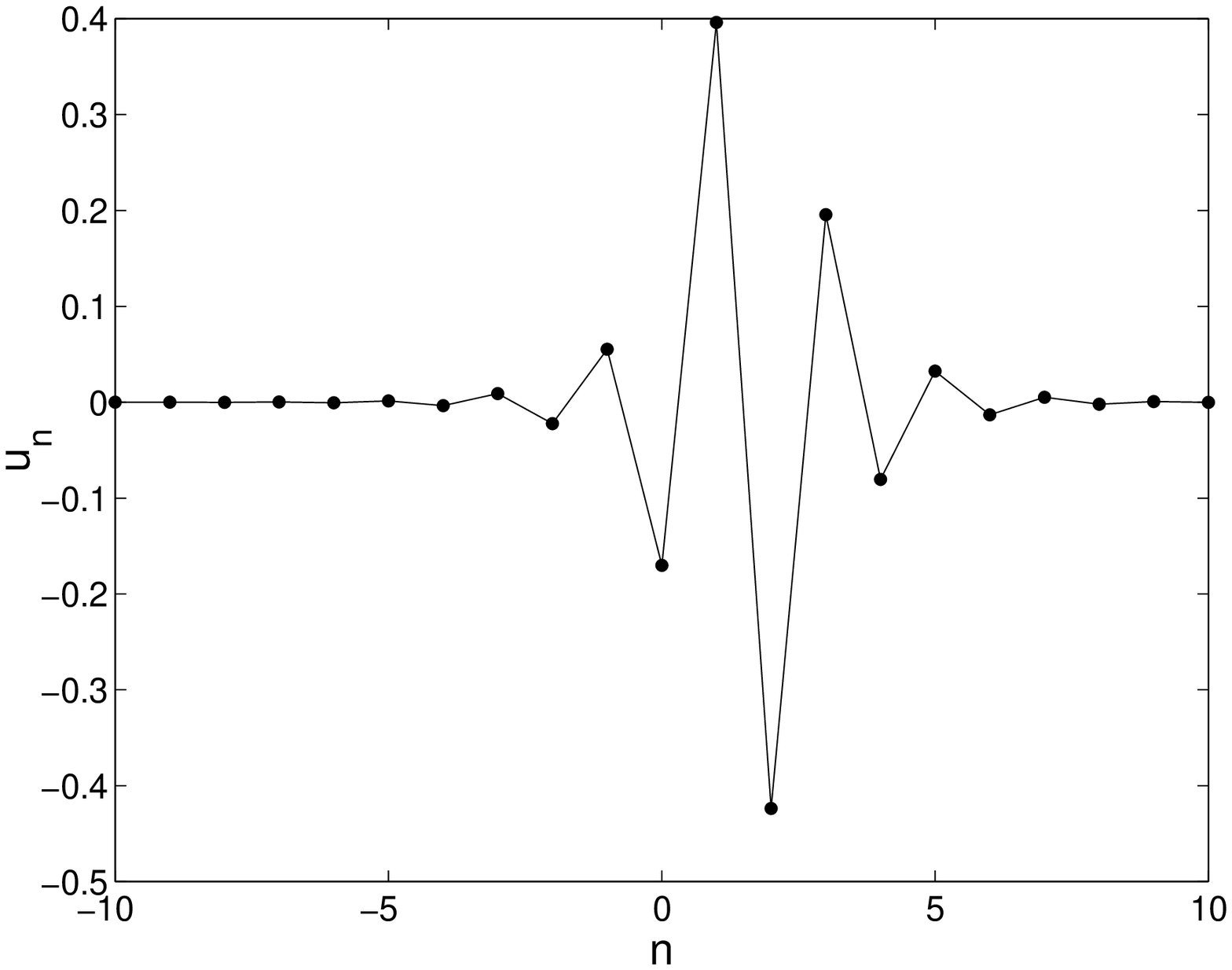}\\
\end{tabular}
\caption{Spatial profiles of the breathers centered at $n=0$ (a),
$n=0.5$ (b), $n=1$ (c) and $n=1.5$ (d) for $\alpha=0.1$. The
results correspond to a chain of oscillators with an on--site
$\phi^4$ potential with breather frequency $\wb=1.2$ and coupling
constant $C=0.09$.}%
\label{fig:brphi4}
\end{center}
\end{figure}

\begin{figure}
\begin{center}
\begin{tabular}{cc}
    (a) & (b) \\
    \includegraphics[width=\middlefig]{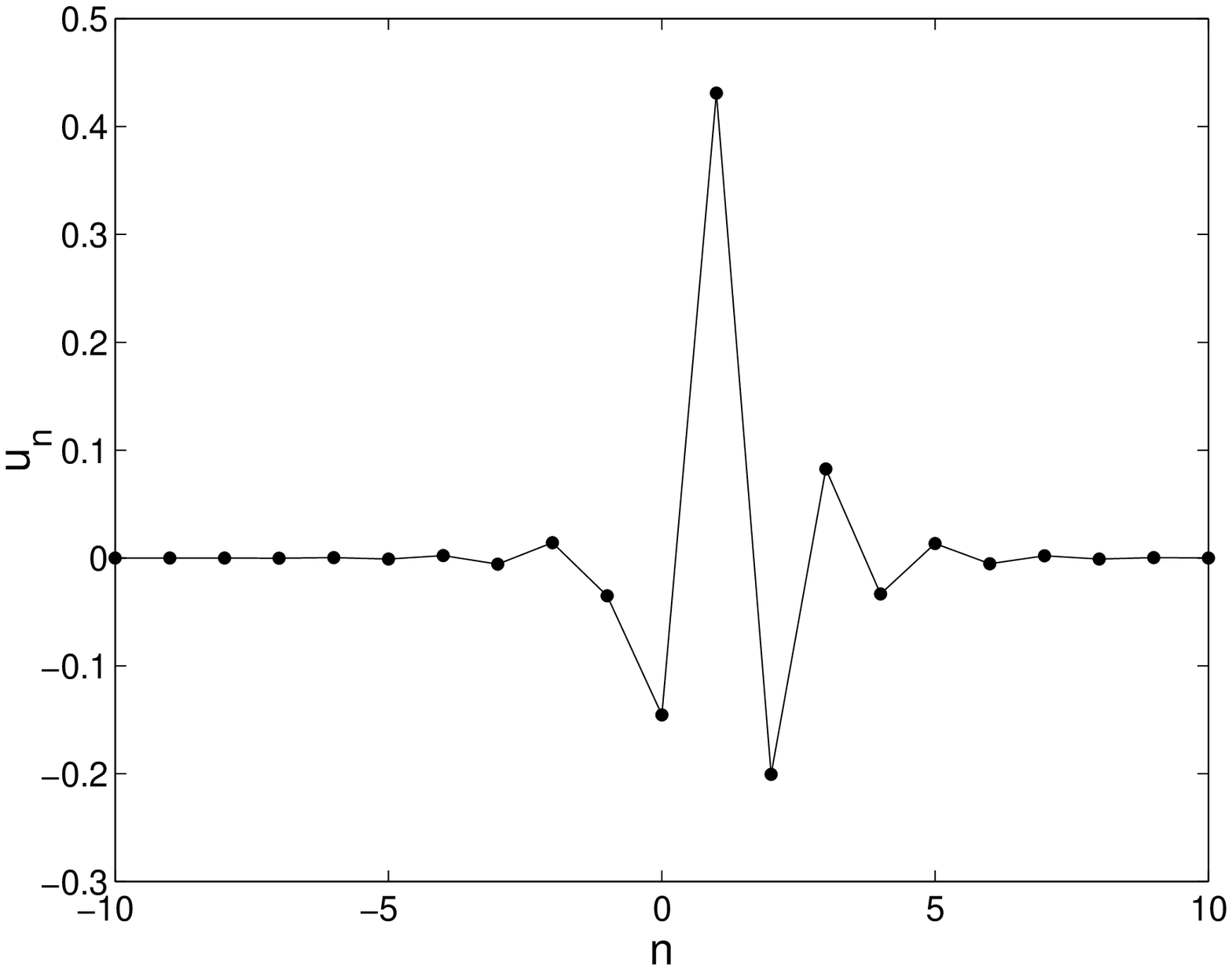} &
    \includegraphics[width=\middlefig]{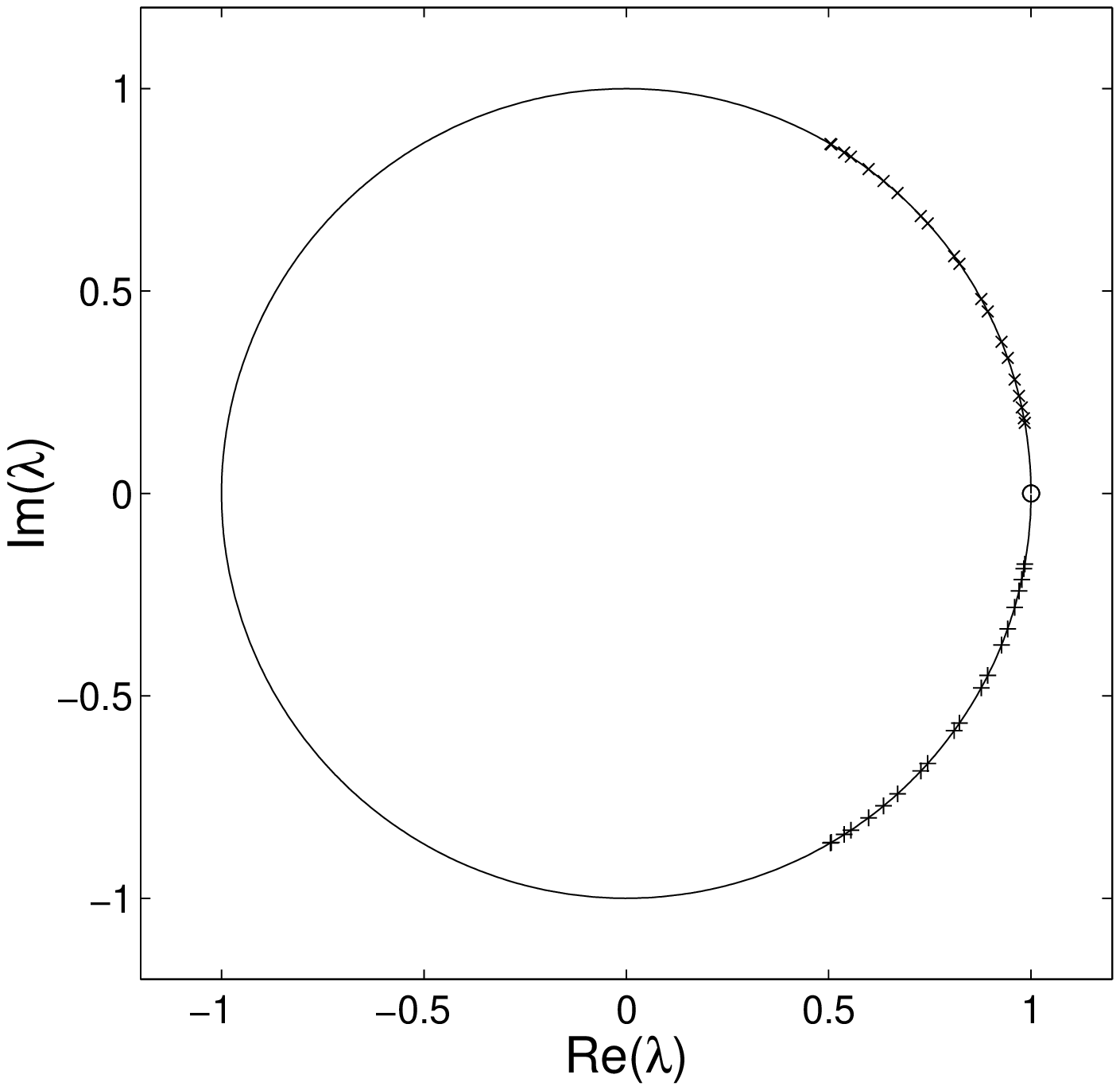}\\
\end{tabular}
\caption{(a) Spatial profiles of the breathers centered at $n=1$
for $\alpha=0.5$ and the same characteristics to those of Figure
\ref{fig:brphi4}. This solution corresponds to the stable ground
state. (b) represents the Floquet eigenvalues of this
solution.}%
\label{fig:stabphi4}
\end{center}
\end{figure}

Some of the breather solutions for $\alpha=0.1$ are shown in
Figure \ref{fig:brphi4} and the profile of the ground state for
$\alpha>0$ together with its Floquet eigenvalues in Figure
\ref{fig:stabphi4}.

\subsection{Instabilities: Switching and Breather Mobility}

\begin{figure}
\begin{center}
\begin{tabular}{cc}
    (a) & (b) \\
    \includegraphics[width=\middlefig]{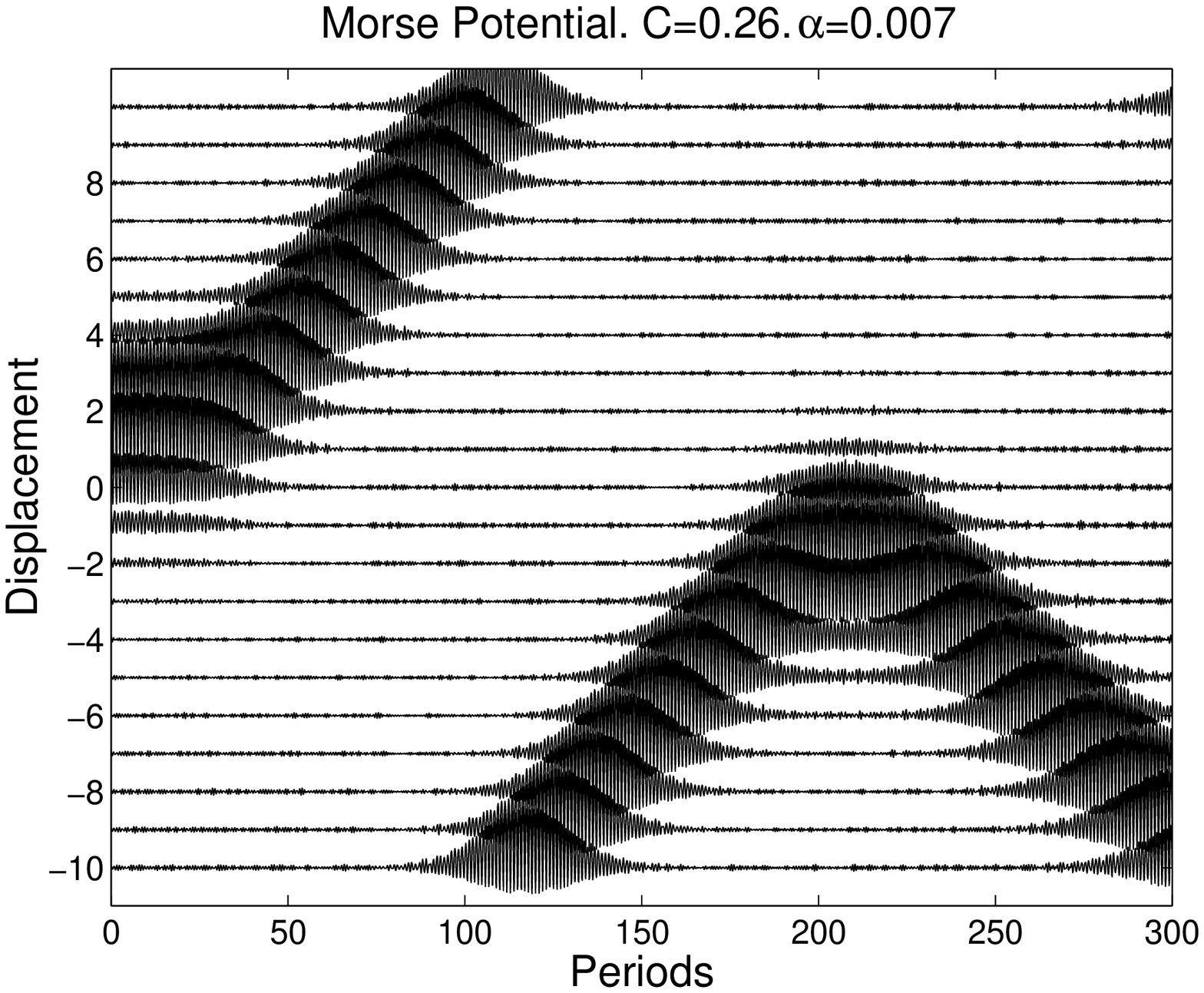} &
    \includegraphics[width=\middlefig]{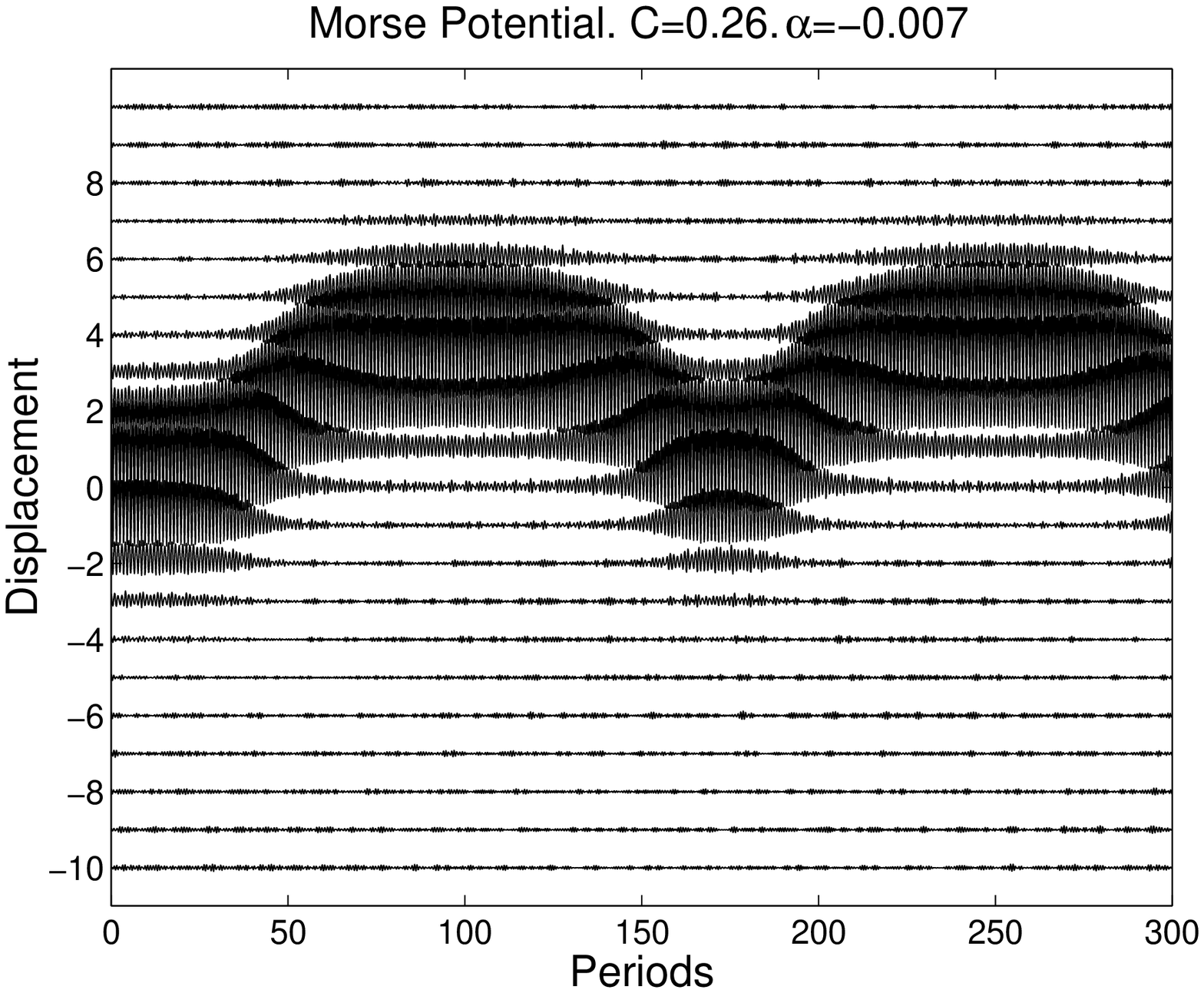}\\
\end{tabular}
\caption{Time evolution of an unstable breather slightly perturbed
for $\alpha>0$ (a) and $\alpha<0$ (b).}%
\label{fig:switchMorse}
\end{center}
\end{figure}

As explained above, the ground state is not always the breather
centered at $n=0$. In particular, for a Morse potential, $C=0.26$,
$\wb=0.8$ and $|\alpha|$ large enough, the ground state for
$\alpha<0$ is the breather centered at $n=1$
(at least in the infinite domain limit, since in the
finite domain case, it may be unstable as shown in Fig.
\ref{fig:stabMorse}).
On the other hand, for $\alpha>0$
the ground state is still the breather at $n=0$. Thus, if an
unstable solution (say, the $n=0$-centered for $\alpha<0$ or the
$n=1.5$ for $\alpha>0$) are perturbed, they are likely to \emph{switch} to
the ground state, in a similar fashion to the phenomenon observed
in \cite{ACMG01}. However, as Figure \ref{fig:switchMorse} shows,
this may not always be the case. In particular,
the breather oscillates between the sites $n=0$ and $n=3.5$
(i.e., in the vicinity of the stable ground state) for
$\alpha<0$, while for $\alpha>0$, it is set in
motion. A possible explanation relies on the fact that the
eigenmode leading to the instability is actually a pinning
mode, that is responsible for breather mobility (see also Section
\ref{sec:dynamic}). Hence a perturbation with the appropriate
sign (i.e., causing the breather to move away from the ground
state of $n=0$ rather than toward it) along this
eigendirection can lead to breather motion.

For a $\phi^4$ potential, $C=0.09$, $\wb=1.2$ and $|\alpha|$ large
enough, the ground state for $\alpha>0$ is the breather centered
at $n=1$, while for $\alpha<0$, it is not well-defined.

\begin{figure}
\begin{center}
    \includegraphics[width=\middlefig]{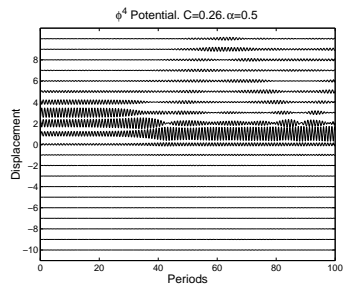}
\caption{Time evolution of an unstable breather slightly perturbed
for $\alpha>0$.}%
\label{fig:switchphi4}
\end{center}
\end{figure}

Figure
\ref{fig:switchphi4} shows the evolution of a 2-site breather
centered at $n=2$ with $\alpha=0.5$.
It is observed that the breather can
switch to the 1-site breather centered at $n=1$, i.e., the ground
state. In this case, no moving localized excitations arise,
but rather the previously suggested switch
occurs. This may be attributed to the fact that for the hard
$\phi^4$ potential,
there do not exist pinning modes that could potentially
lead to breather mobility (see also \cite{CAT96}).

\section{Dynamic Results: Moving Breathers and Breather-Bend Interactions}
\label{sec:dynamic}

In this section we examine how the local geometry of the bend can
influence the motion of a breather in its vicinity. In particular,
the moving breathers are launched towards the bend vertex
following the marginal mode method \cite{CAT96}. The latter
consists of adding to the velocities of the static breather a
perturbation of magnitude $\lambda$, collinear to the direction of
a specific linear localized mode, namely the eigenmode
corresponding to translation. Then, the kinetic (translational)
energy of the breather can be defined as $K=\lambda^2/2$. As Ref.
\cite{CPAR02} shows, the moving breather can be described as a
quasi-particle of mass $\m$ defined through the relation
$K=\lambda^2/2=\m v^2/2$, with $v$ being the translational
velocity of the breather. The mass is a measure of the inertia of
the breather to external forces.

Another useful concept for our study is the breather
energy center, which is a measure of the position of the
localized structure. It is defined through the relation $X=(\sum_nne_n)/E$,
i.e., as the center of mass with respect
to $e_n$, the energy density at the n-th site.
$E=\sum_ne_n$ is the total energy of the breather \cite{CAGR02}.

The local ``coupling inhomogeneity'' induced by the bend, should
result in features similar to those observed in Ref.
\cite{CPAR02b}. In particular, for a soft potential, if
$\alpha>0$, the bending should act, in principle (see also below), as a potential barrier, i.e.,
the breather can cross the vertex as long as its translational
energy is above a threshold value. If this condition is satisfied,
the breather decelerates when crossing the bend, but eventually
recovers its initial velocity. If, on the other hand, $\alpha <
0$, then the corresponding behavior is more complex: for small
values of $|\alpha|$, the bend acts as a potential well,
accelerating a moving breather approaching it. However, for larger
values of the parameter, trapping and even reflection become
possible. For a hard potential, such as $\phi^4$, the behaviors
for $\alpha>0$ and $\alpha<0$ are reversed \cite{CPAR03}.

We will focus on the $\alpha>0$ case and consider, as before, a
chain of oscillators with 1) the Morse on--site potential, as well
as one with 2) the $\phi^4$ on--site and inter--site potential.
Thus, for the case of the Morse (soft) potential, the bending will
act as a potential barrier, while for the $\phi^4$ (hard)
potential, the well/trapping/reflection regimes will be examined.

\subsection{Morse Chain: Bend-Induced Potential Barrier}

\begin{figure}
\begin{center}
\begin{tabular}{cc}
    (a) & (b) \\
    \includegraphics[width=\middlefig]{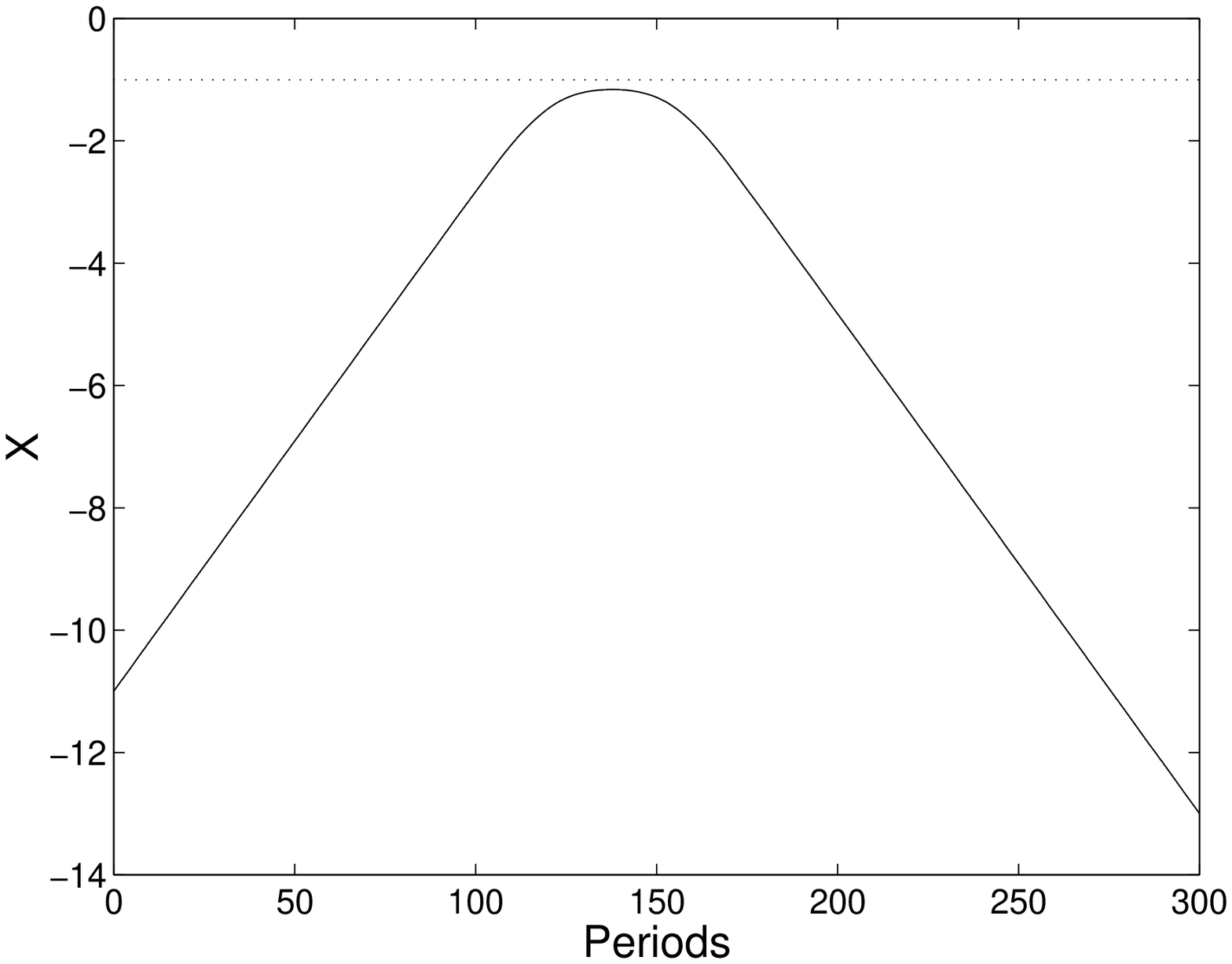} &
    \includegraphics[width=\middlefig]{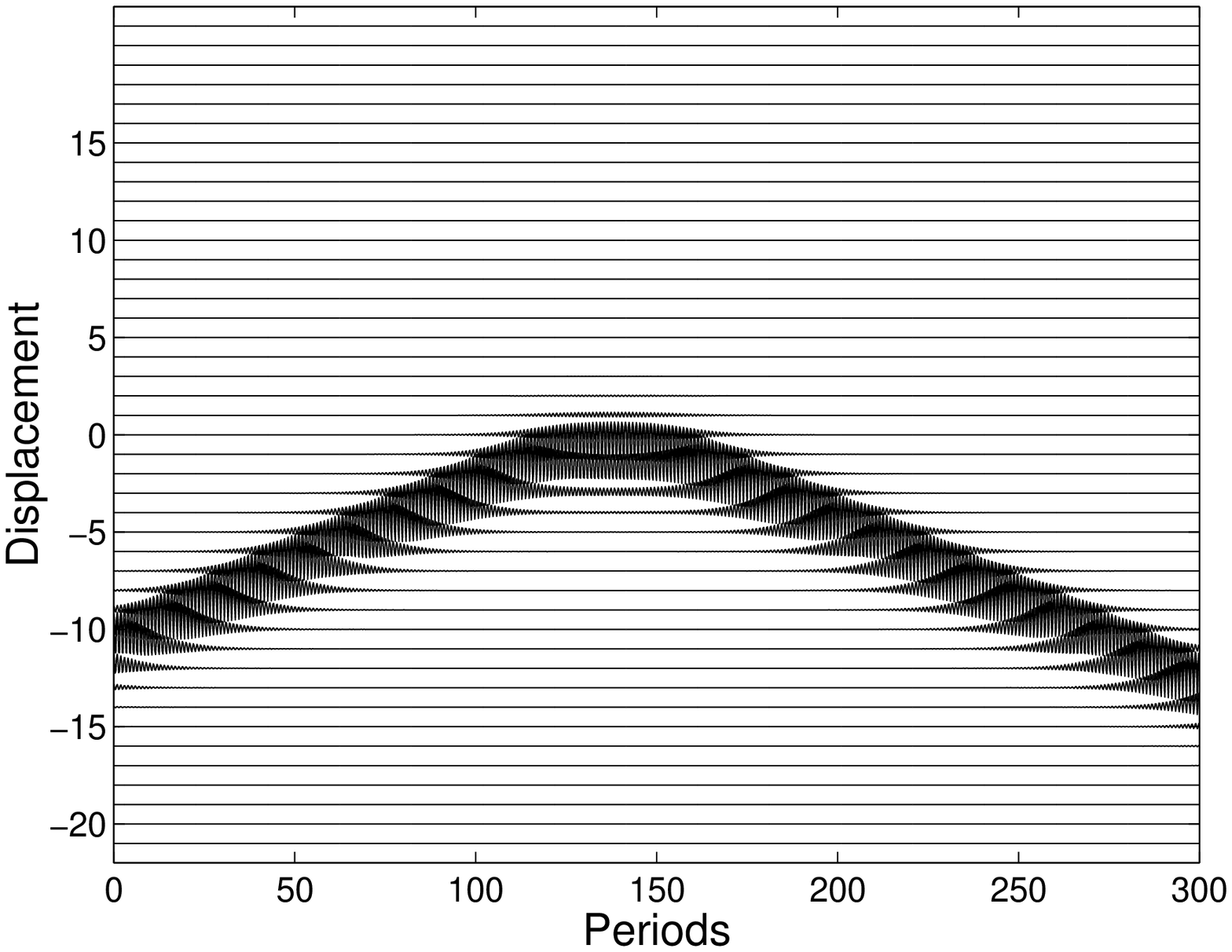}\\
    (c) & (d) \\
    \includegraphics[width=\middlefig]{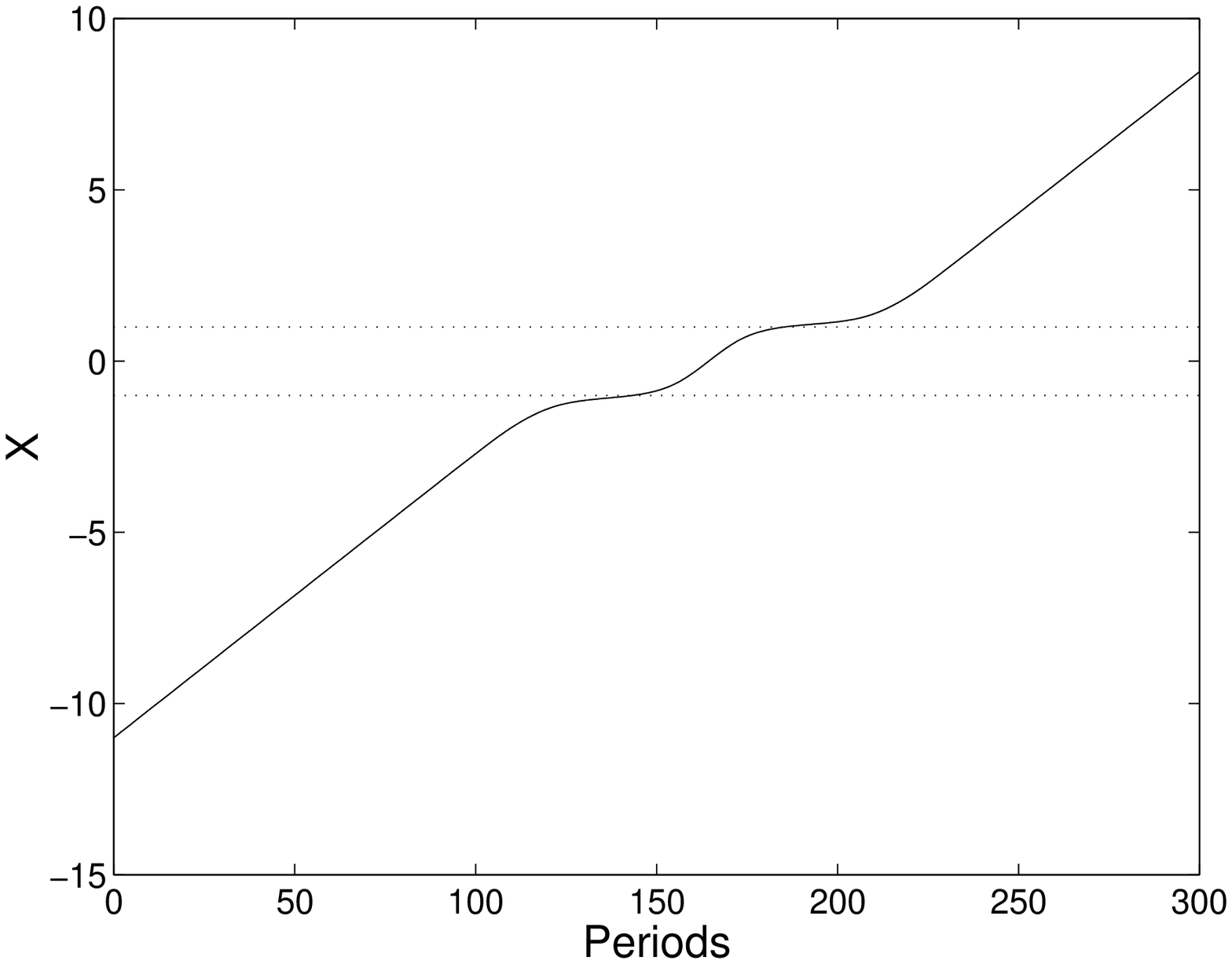} &
    \includegraphics[width=\middlefig]{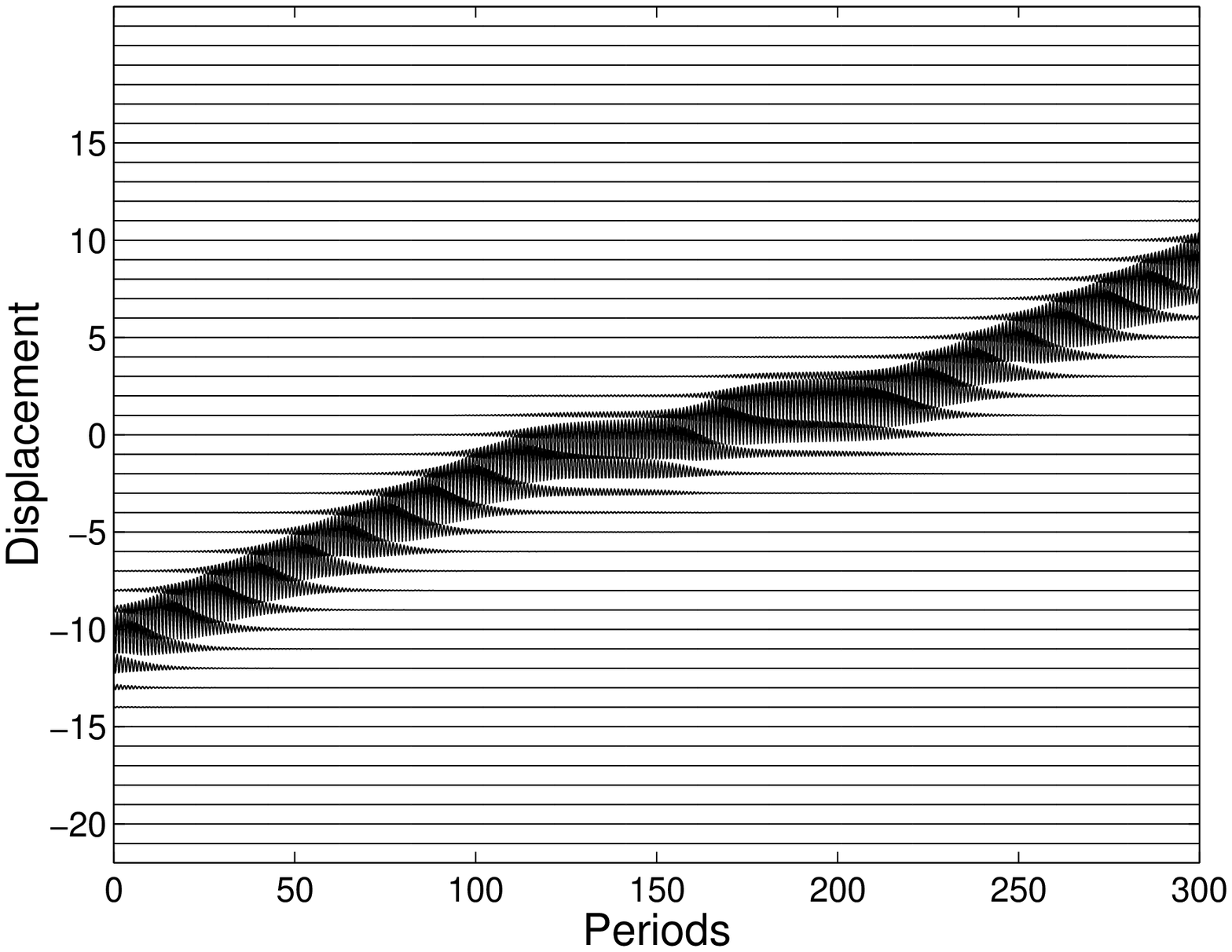}\\
    (e) & (f) \\
    \includegraphics[width=\middlefig]{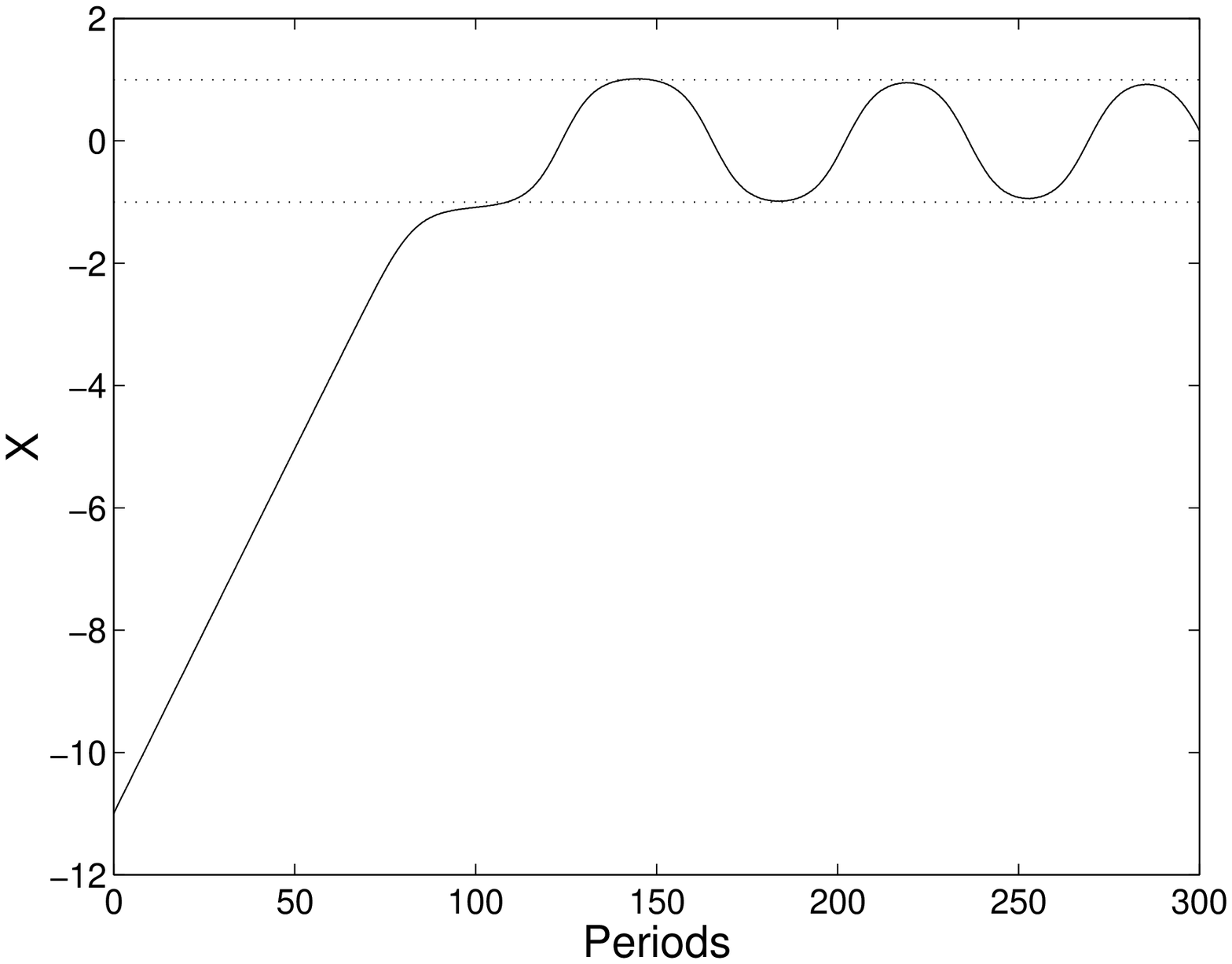} &
    \includegraphics[width=\middlefig]{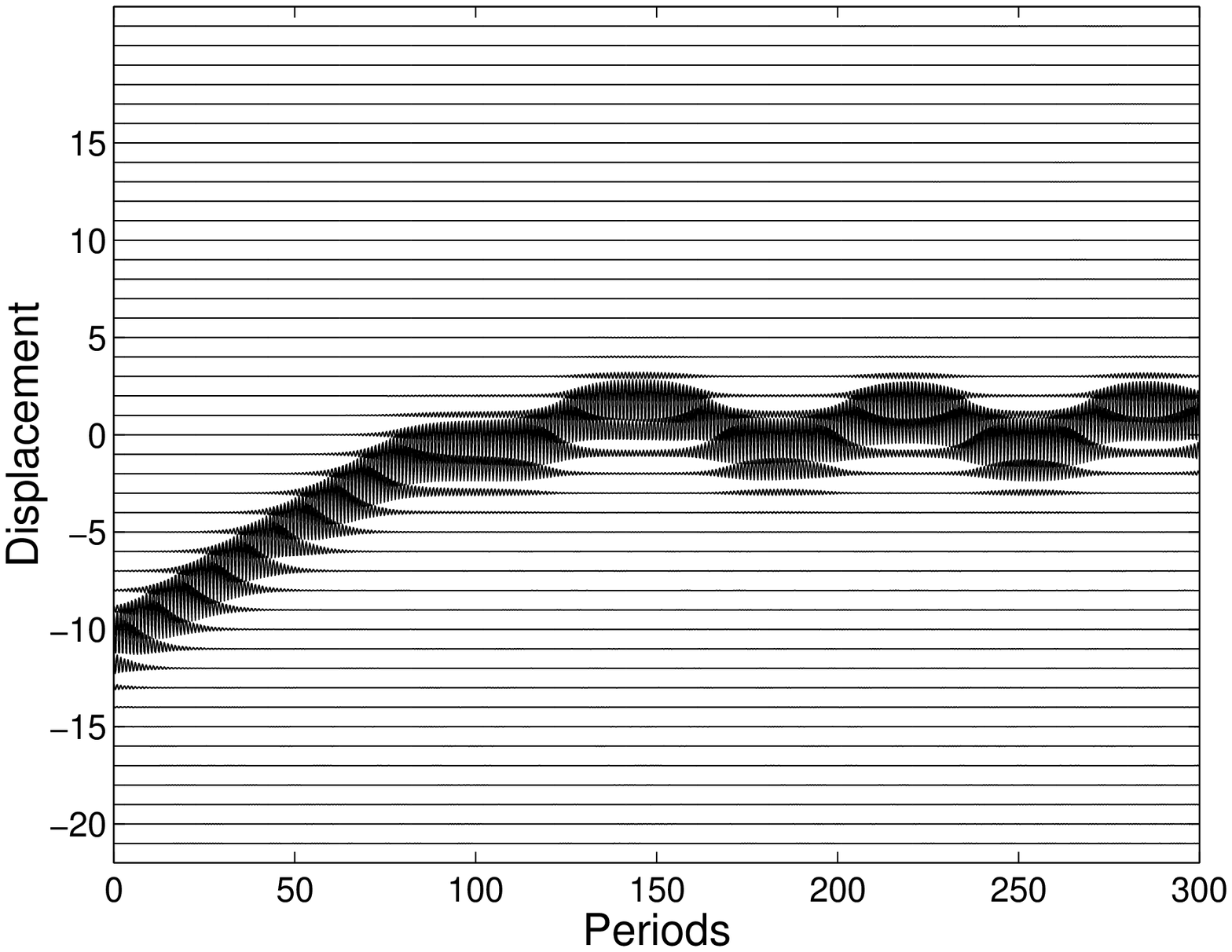}\\
\end{tabular}
\caption{Evolution of the energy center (left) and the moving
breather (right) for a Morse on--site potential with parameters
$\wb=0.8$ and $C=0.13$. Figures (a) and (b) correspond to a
reflection ($\alpha=0.008$ and $K=0.00174$), (c) and (d) to a
refraction ($\alpha=0.008$ and $K=0.00180$) and (e) and (f) to a
refraction through the site $n=-1$ and a reflection at $n=1$
leading to a trapping ($\alpha=0.008$ and $K=0.00392$).}%
\label{fig:moving_barrier}
\end{center}
\end{figure}

As mentioned earlier, for a soft on--site potential and
$\alpha>0$, the bending acts as a potential barrier, i.e., if the
translational energy of the breather $K$ is above a threshold
value $U_c$, the breather crosses the bending. Otherwise, the
breather is reflected. Figure \ref{fig:moving_barrier} shows the
time evolution of a reflected (top panel), a refracted (middle panel)
and that of a trapped
breather (bottom panel). In the top panel (for $K<U_c$),
the breather is immediately
reflected from the potential barrier, emerging with (approximately) the
same speed, but propagating in the opposite direction.
In the middle panel of the figure, it can be
observed that, for $K>U_c$ the breather decelerates when it
approaches the $n=-1$ site, accelerates when it crosses this point
until it reaches the bend vertex ($n=0$ site); it decelerates again
when approaching the $n=1$ site, accelerating when it crosses this
point.

\begin{figure}
\begin{center}
    \includegraphics[width=\singlefig]{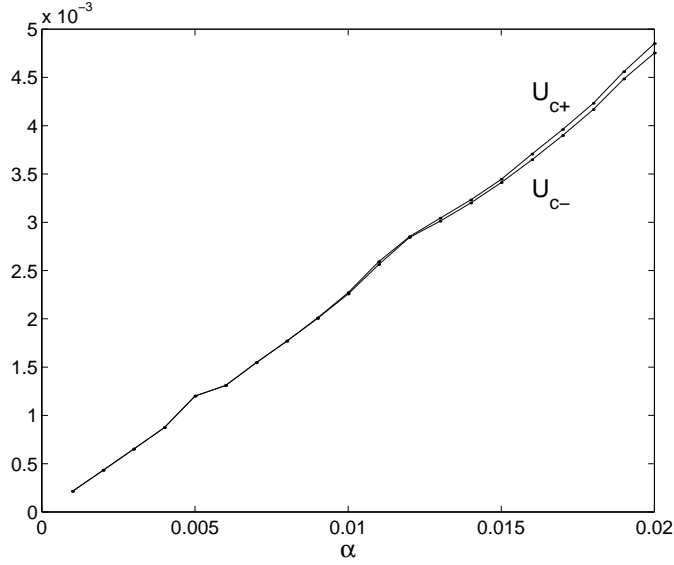}
\caption{Dependence of the critical energy $U_c$ on
$\alpha$ for a  Morse on--site potential with parameters $\wb=0.8$
and $C=0.13$. Note that for sufficiently large $\alpha$, the curve
separates in two parts corresponding to $U_{c+}$ and $U_{c-}$ (see also the
relevant discussion in the text).}%
\label{fig:lamc_barrier}
\end{center}
\end{figure}

Figure \ref{fig:lamc_barrier} shows the critical energy $U_c$
necessary to cross the bend. An important point to indicate is the
existence of two {\it distinct} values of the critical energy
$U_{c\pm}$, that correspond respectively to the kinetic energy
threshold for crossing the site $n=\pm1$. When $U_{c-}<U_{c+}$
holds, for intermediate values of $U_{-} \leq K < U_+$,  the loss
of energy due to phonon radiation in the breather-bend
interaction, disallows the moving breather from crossing the bend.
Hence, if the two thresholds are different, the breather can be
trapped oscillating between the $n=-1$ and $n=1$ sites; an example
is shown in the bottom panel of Fig. \ref{fig:moving_barrier}. As
is naturally expected, the energy loss decreases when $\alpha$ is
small (i.e., as we approach the rectilinear, integer-shift
invariant chain), and $U_{c-}=U_{c+}$ is practically fulfilled in
this parameter range, e.g., for $\alpha <0.01$ in Fig.
\ref{fig:lamc_barrier}.

The potential barrier $U=U(X)$ can be calculated using the
procedure described in \cite{CPAR02}: if $K<U_c$, the value of $K$
is fixed and $X_o$ is the point where the breather is reflected
(turning point), then $K= U(X_o)$.

If, on the other hand, $K>U_c$, the barrier can be calculated
through the formula:

\begin{equation} \label{eq:barrier2}
    U(X)=K\left(1-\left(\frac{v(X)}{v_o}\right)^2\right),
\end{equation}%
where $v_o$ is the initial translational velocity of the breather
and $v(X)$ is its (numerically computed) velocity at the point
$X$. If the mass of the breather is constant, the barrier
calculated for $K<U_c$ and $K>U_c$ should coincide, except for the
oscillations that appear in the second case with respect to the
first one. The origin of these oscillations relies on the
non-uniformity of the instantaneous translational velocity of the
breather due to the discreteness of the system. Thus, to obtain
the actual shape of the barrier, the turning point approach should
be used.

However, this method can only be applied in our case for $X<-1$
(and hence by symmetry for $X>1$) as for $K<U_c$, the breather
cannot cross over to the region $-1<X<1$. In order to overcome
this drawback, we have approximated the potential barrier created
by a single inhomogeneity with a Gaussian shape:
$U(X)=a\exp(-b(X-X_o)^2)$, where $X_o$ is the point where the
maximum of the barrier is located (which need not be the
inhomogeneity point) \cite{pgk1,Cuevas,pgk3}. In our bend setting,
there are two points where the equation is inhomogeneous with
respect to the rest of the chain. Hence, it is natural to
approximate the actual barrier with the linear superposition of
two Gaussian barriers, i.e.,

\begin{equation} \label{eq:barrier1}
    U(X)=a\exp(-b(X-X_{-1})^2)+\exp(-b(X-X_1)^2), \qquad
    x_{\pm1}=\pm(1+\delta),
\end{equation}%
with $\delta$ being the difference between the turning point at
$K\approx U_c$ and the ``impurity site'';
e.g. for a breather launched from the left of the bending,
$\delta=X(U_c)+1$. Thus, equation (\ref{eq:barrier1}) becomes:

\begin{equation} \label{eq:barrier}
    U(X)=2a\exp(-bX^2)\exp(-b(1+\delta)^2)\cosh(2bX(1+\delta)).
\end{equation}

\begin{figure}
\begin{center}
    \includegraphics[width=\singlefig]{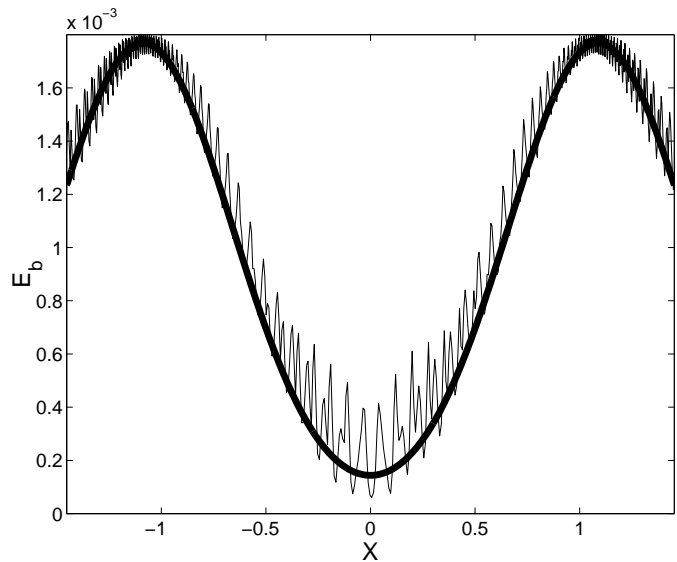}\\
\caption{Potential barrier experienced by the moving breather
in the chain with a Morse on--site potential, when it
reaches the bending point.
Parameters of the breather are $\wb=0.8$, $C=0.13$ and
$\gamma=0.08$. The thick line corresponds to the fit to equation
(\ref{eq:barrier}) and the slim line to equation
(\ref{eq:barrier2}) for $K=0.0018$. The oscillations of the last
curve are due to the non-uniformity of the translational velocity
of the breather. Parameters of the fitting to equation
(\ref{eq:barrier}) are: $a=0.001768$, $b=2.7068$ and
$\delta=0.08754$.}%
\label{fig:barrier}
\end{center}
\end{figure}

Figure \ref{fig:barrier} shows the barrier calculated using
equations  (\ref{eq:barrier2}) and (\ref{eq:barrier}). Parameters
$a$, $b$ and $\delta$ are chosen through the Gaussian fit of the
barrier points for $|X|>1$. Comparing the fitted profile of
Eq. (\ref{eq:barrier}) with the one obtained for a supercritical
case of $K>U_c$ indicates very good agreement, which in turn supports
the assumption of Eq.  (\ref{eq:barrier1}).

Notice, however, that the such a restricted one-degree-of-freedom
picture cannot capture phenomena such as emission of radiative
small amplitude waves during the crossing of the bend, and
consequently cannot display a phenomenology as rich as the original
problem (e.g., it cannot exhibit the trapping type phenomena discussed above).

\subsection{$\phi^4$ Chain: Well/Trapping/Reflection Regimes}

\begin{figure}
\begin{center}
\begin{tabular}{cc}
    (a) & (b) \\
    \includegraphics[width=\middlefig]{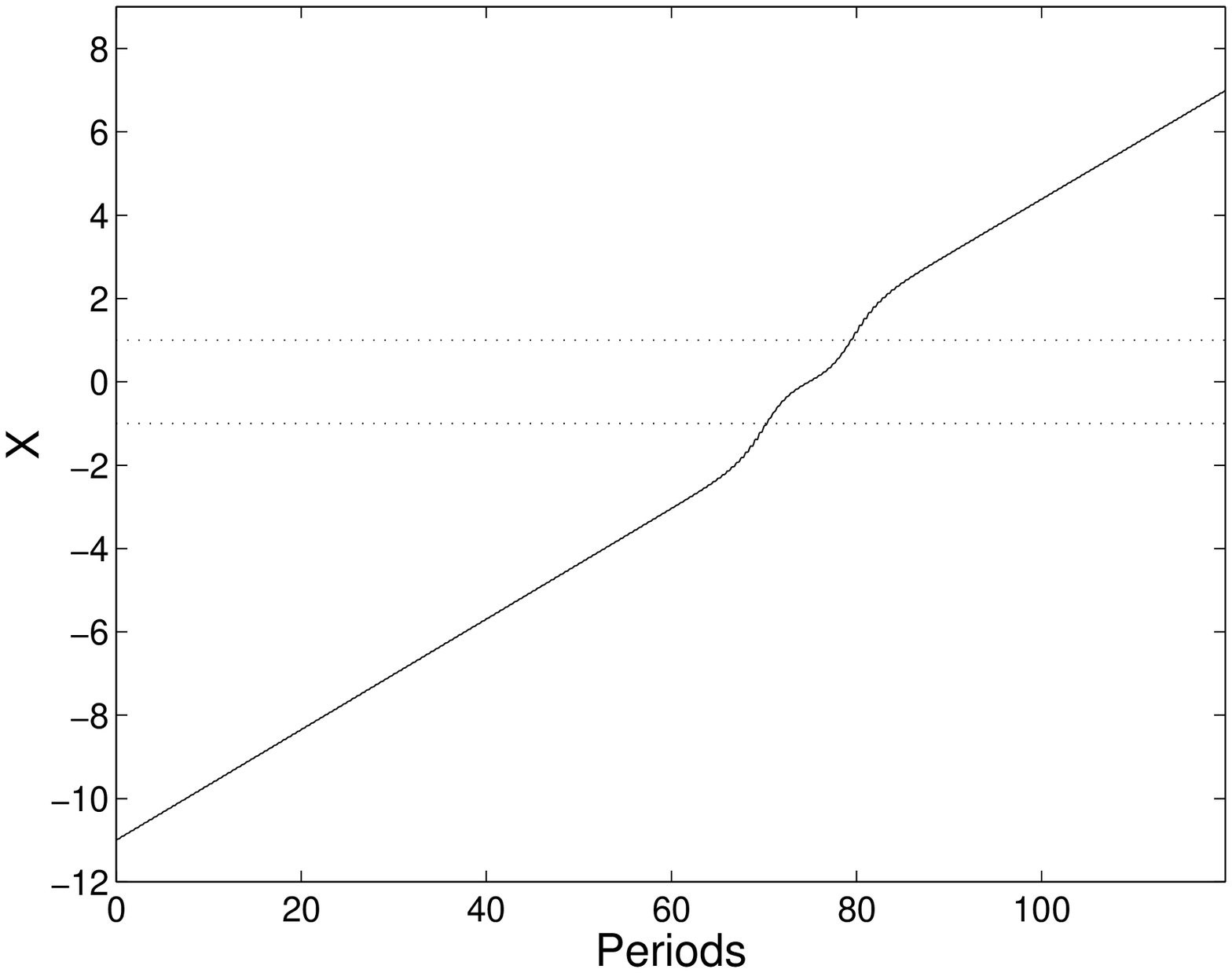} &
    \includegraphics[width=\middlefig]{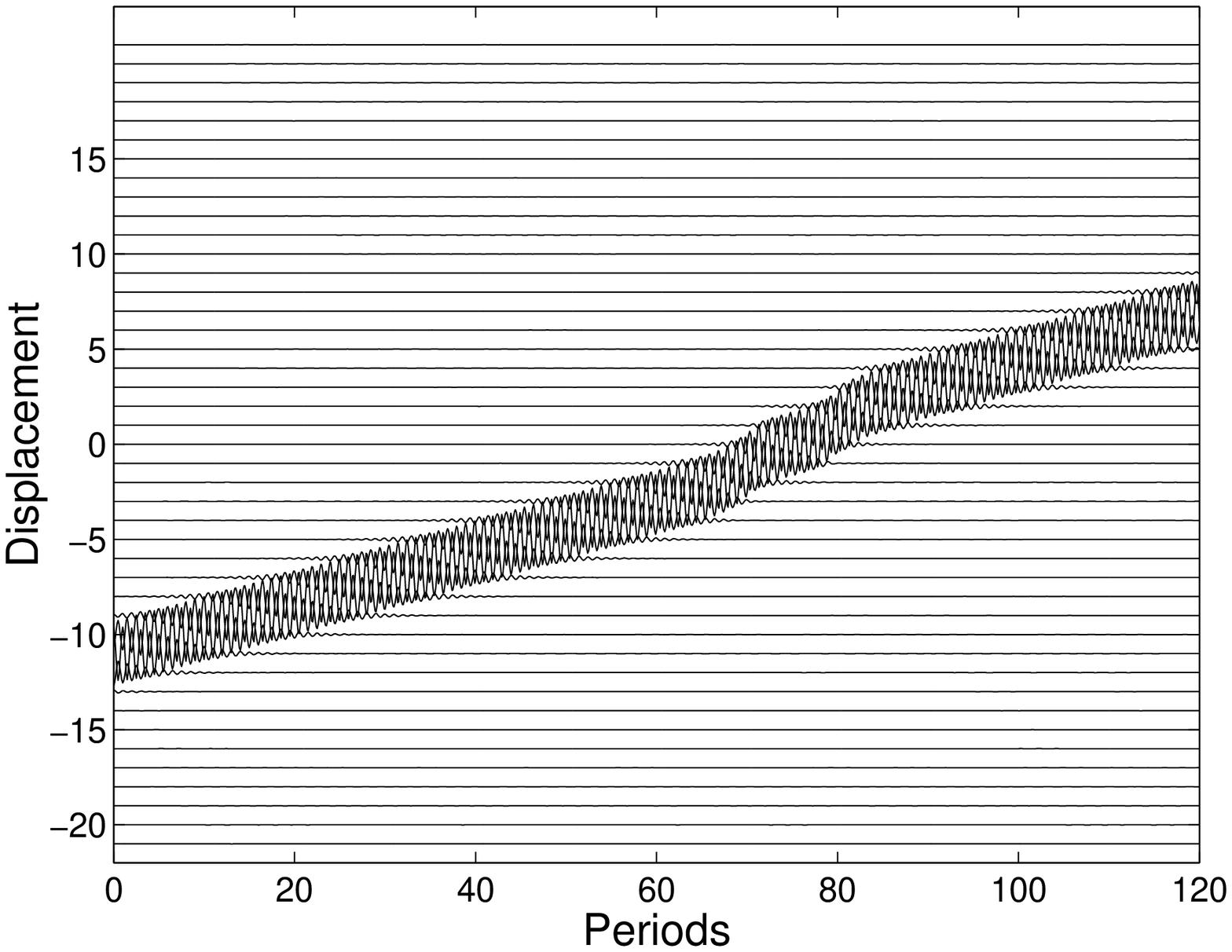}\\
    (c) & (d) \\
    \includegraphics[width=\middlefig]{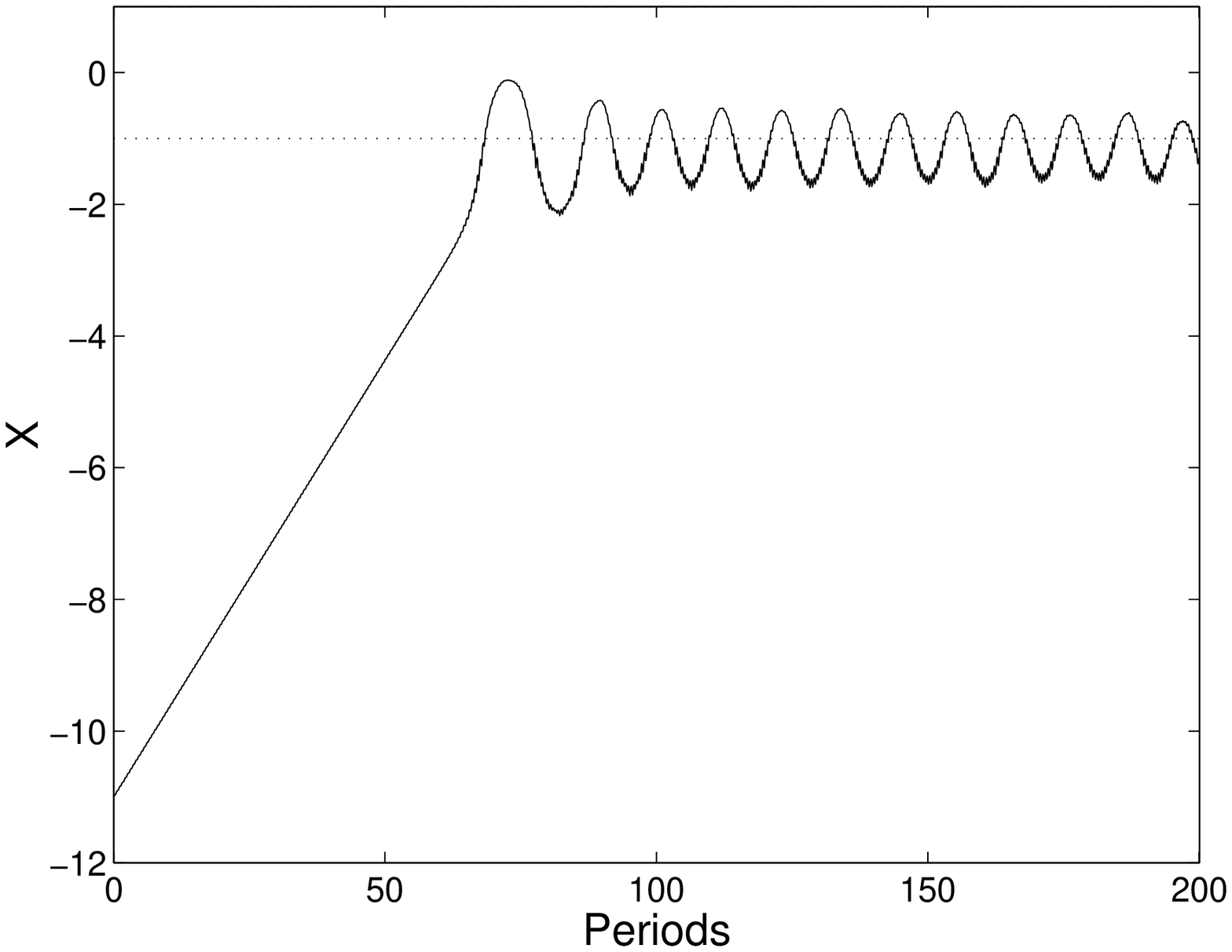} &
    \includegraphics[width=\middlefig]{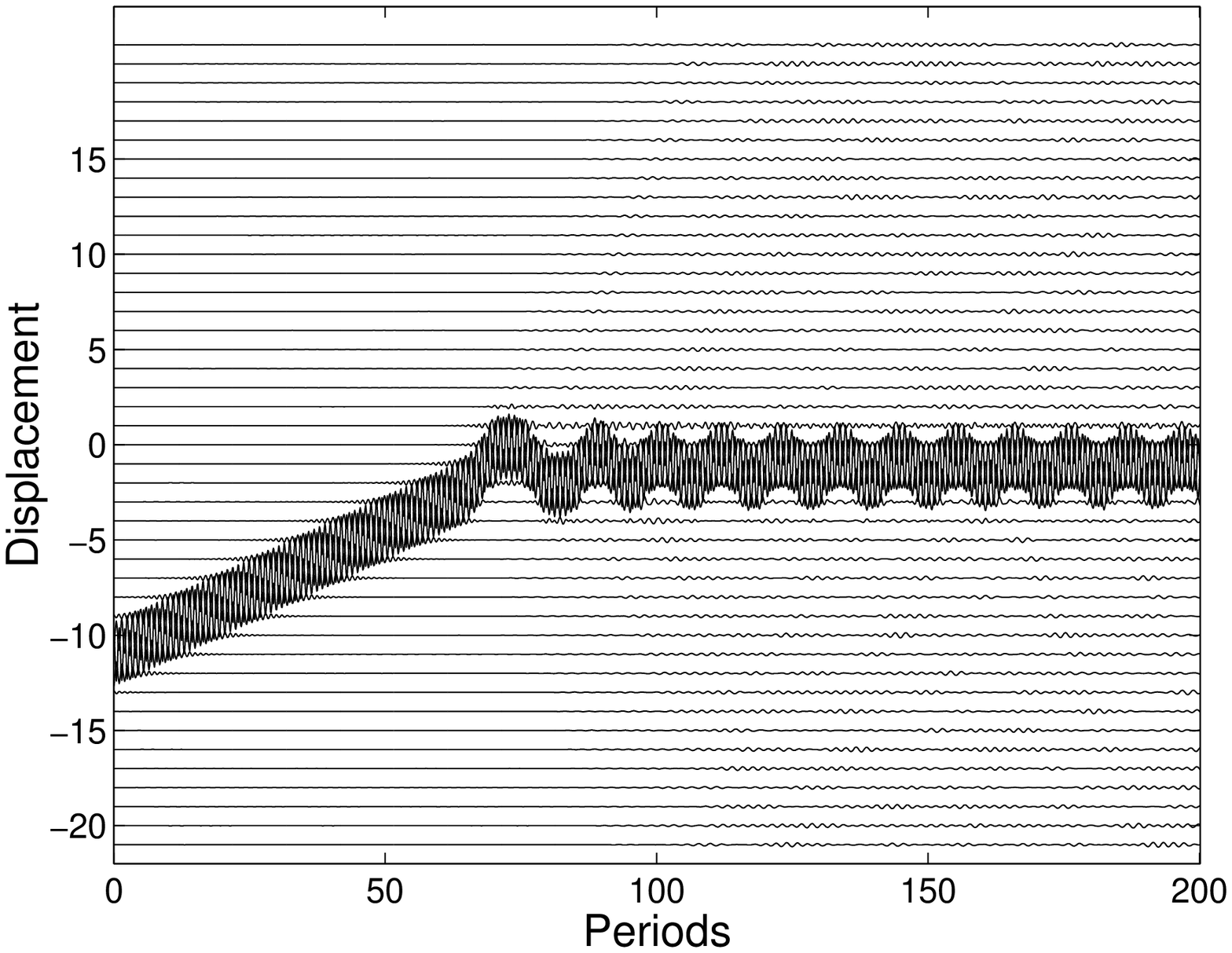}\\
    (e) & (f) \\
    \includegraphics[width=\middlefig]{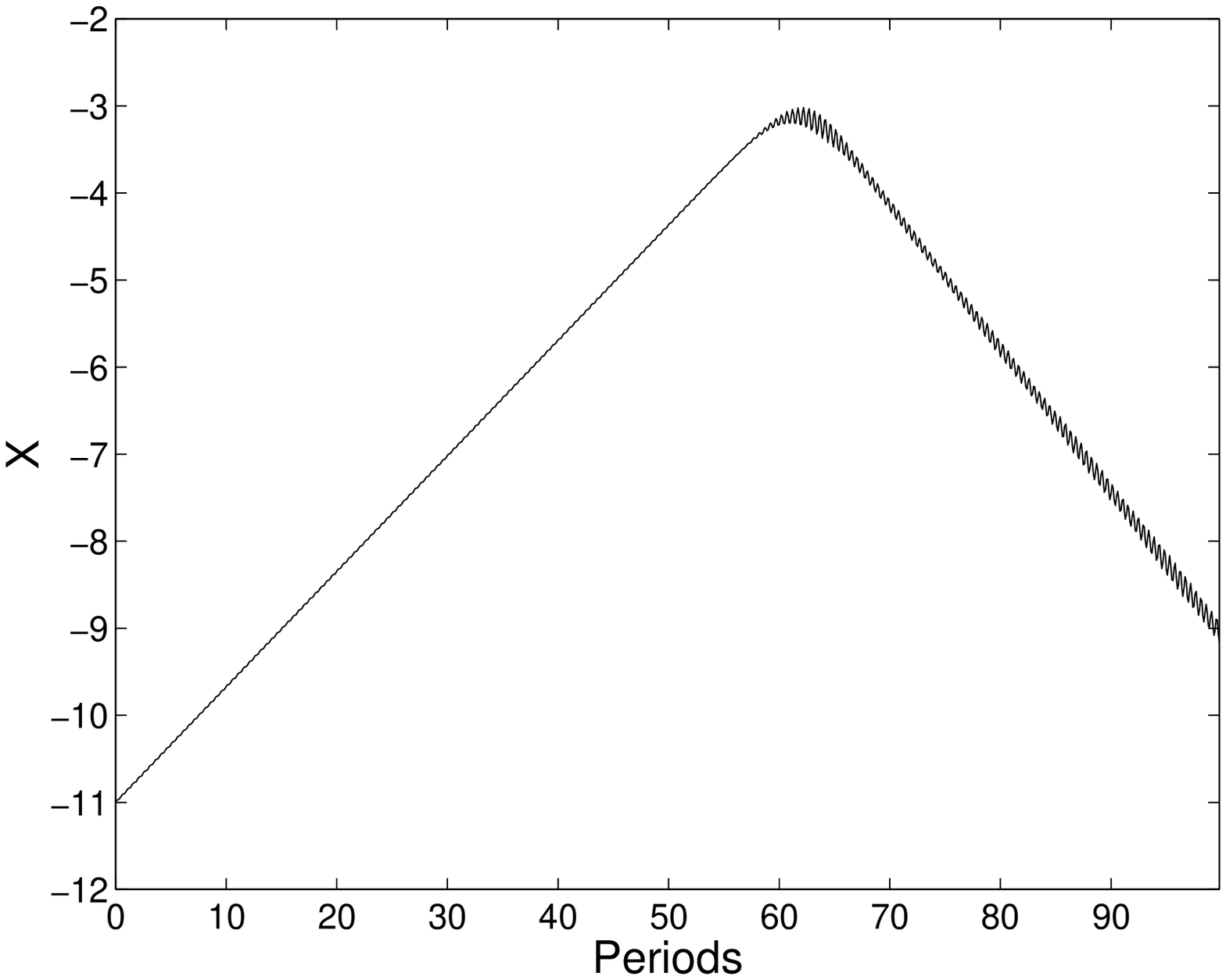} &
    \includegraphics[width=\middlefig]{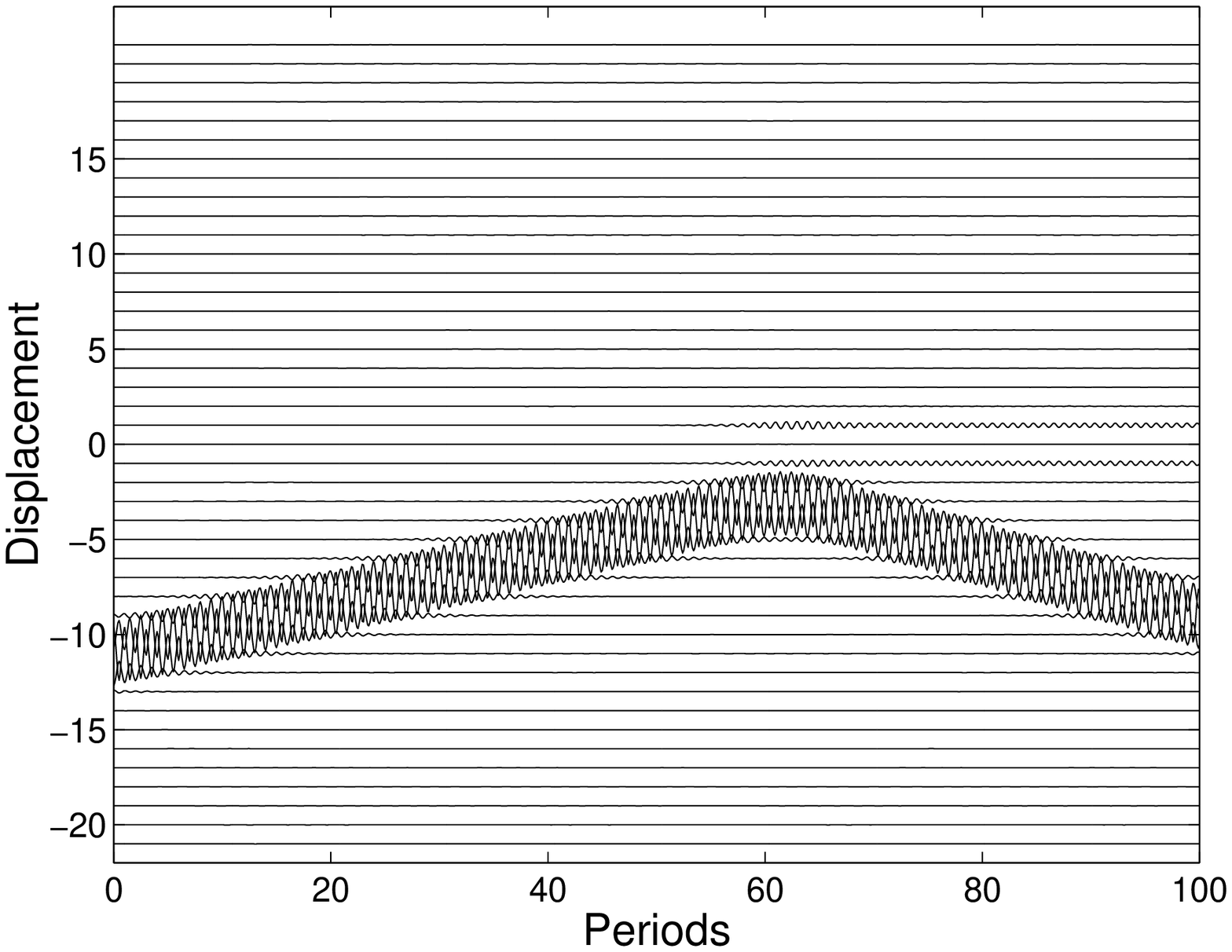}\\
\end{tabular}
\caption{Evolution of the energy center (left) and the moving
breather (right)  for $\phi^4$ on--site and interaction potentials
with parameters $\wb=3$ and $C=0.5352$. Figures (a) and (b)
correspond to a refraction ($\alpha=0.2$), (c) and (d) to a
trapping ($\alpha=0.5$) and (e) and (f) to a reflection ($\alpha=4$).}%
\label{fig:moving_trapping}
\end{center}
\end{figure}

For the case of a hard potential with $\alpha>0$ three different
behaviors can be observed when $\alpha$ increases (see Figure
\ref{fig:moving_trapping}):
\begin{enumerate}
\item For small values of $\alpha$, the breather accelerates when
reaching the bend points, and eventually emerges from the bend
recovering a constant velocity (but not the same as the initial one
due to radiative emissions). That is, the bending acts as a
potential well.

\item For larger values of $\alpha$, the breather is trapped at
the $n=-1$ site.

\item For even larger values of $\alpha$, the breather is
reflected and the particle at the vertex remains excited. This
phenomenon can be also called partial trapping, since a fraction
of the original breather energy remains localized at the bend.

\item Finally, beyond a third critical point for $\alpha$, the
breather is reflected from the bend, and, in a different fashion
to the barrier case, the breather is not refracted for any
velocity.
\end{enumerate}

\begin{figure}
\begin{center}
\begin{tabular}{cc}
    (a) & (b) \\
    \includegraphics[width=\middlefig]{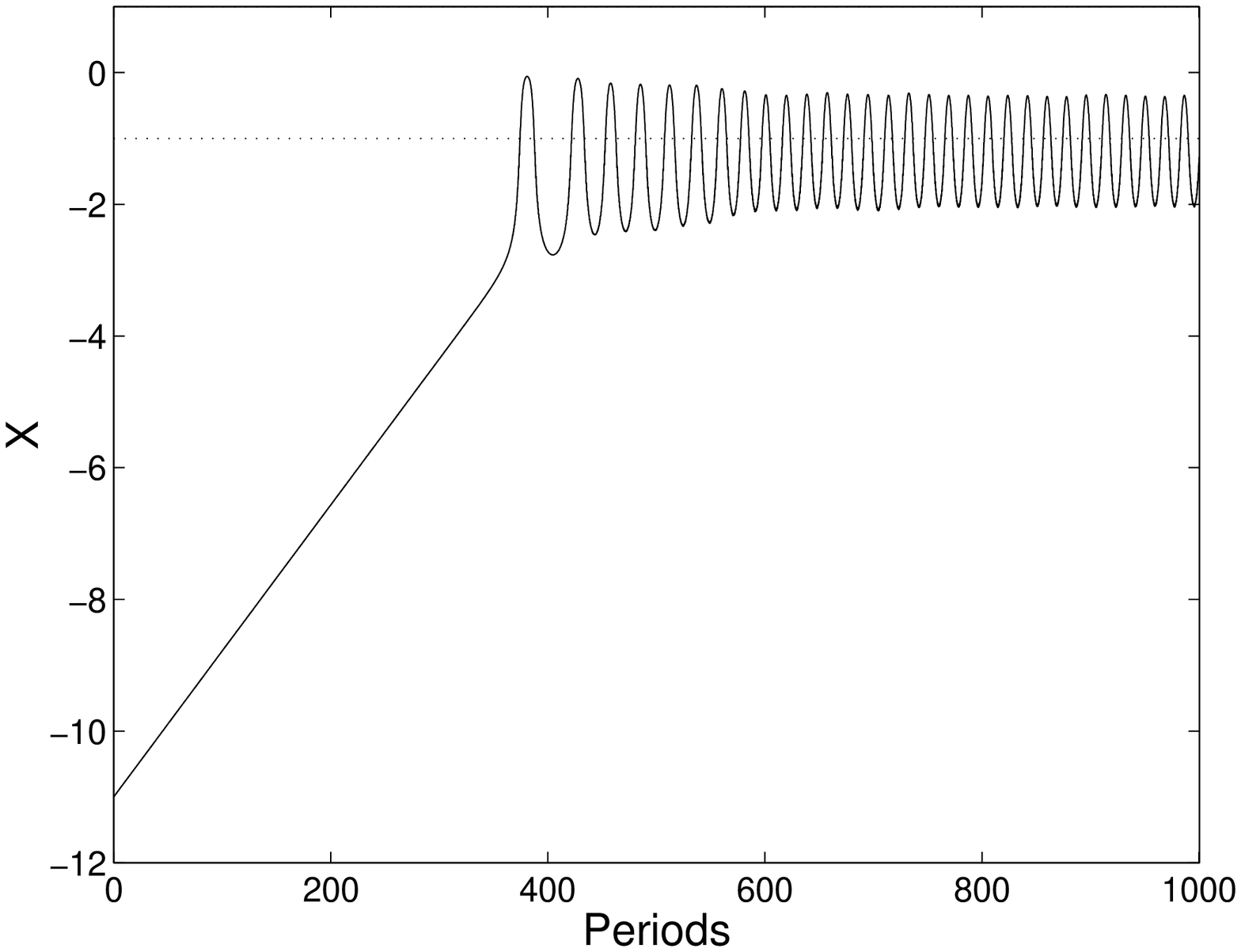} &
    \includegraphics[width=\middlefig]{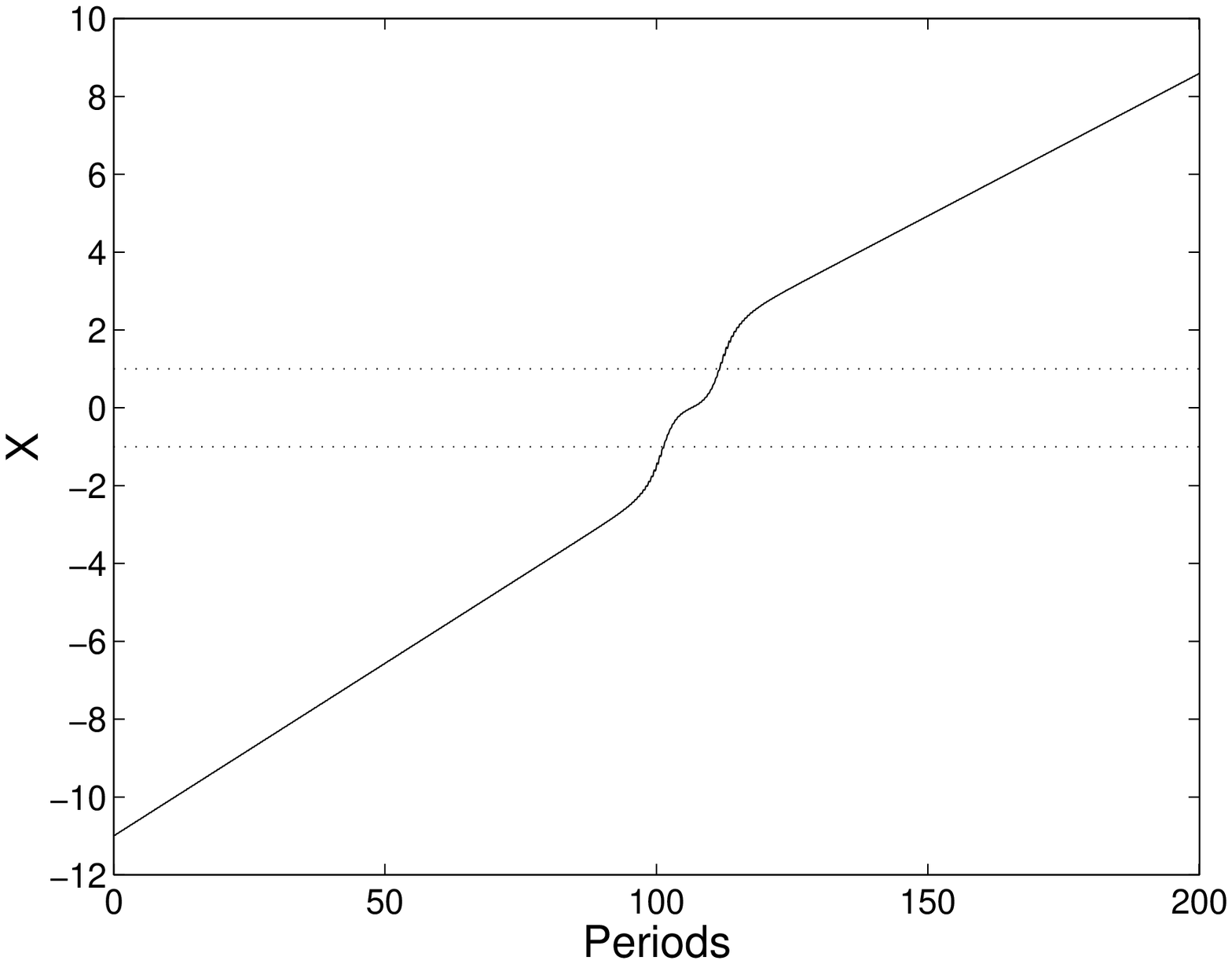}\\
\end{tabular}
\caption{Evolution of the energy center for a $\phi^4$ on--site
and interaction potentials with parameters $\wb=3$, $C=0.5352$ and
a fixed value of $\alpha=0.3$. (a) corresponds to a trapping case
($K=0.00125$) and (b) to a well regime ($K=0.02$).}%
\label{fig:moving_trapping2}
\end{center}
\end{figure}

It should be noted that the boundaries between partial trapping to
reflection and well to trapping regimes are not well-defined. In
the first case, the amplitude of the vertex excitation decreases
when $|\alpha|$ increases. Due to this fact, it is difficult to
establish the borderline between those regimes. In the second
case, the breather is trapped if its translational energy is below
a threshold value. However, for energies above this value, the
bend acts as a potential well. Thus, the trapping takes place
without need of energy loss when the breather crosses the $n=-1$
site, that is, it occurs as long as the translational energy is
below the threshold commented above. These facts are illustrated
in Figure \ref{fig:moving_trapping2}.

\section{Conclusions}

In this communication, we have introduced a discrete Klein-Gordon chain
model variant, emulating the existence of a geometric bend at a lattice
site.

We have examined the static properties of the localized in space, time
periodic solutions that exist in this setting, starting from the limit
where the chain is rectilinear. In the latter case, from
the integer shift invariance and numerous earlier studies \cite{review},
we know that site-centered and bond-centered solutions will exist and
their stability can be established. Hence, using continuation from
the ``straight'' limit into the ``bent'' regime, we are able to trace
the branches of corresponding solutions and identify their saddle-node
bifurcations leading to the termination of the corresponding branches.
However, we have also identified symmetry breaking effects in which one
of the asymmetric branches does {\it not} collide with one of its
neighboring counterparts to terminate its existence, but rather survives
throughout the bend-parameter strength continuation,
and becomes the lowest energy state of the lattice. Excitation of
unstable bend modes in such cases will lead to trapping
around (i.e., {\it switching to}) such
an asymmetric state, but in the presence of pinning modes, moving
breathers may result.
Hence, we conclude
that sufficiently strong bends may favor an asymmetric localization of energy,
with respect to the bend center.

We have also studied the dynamic properties of the interaction of such
bends with moving localized modes, when the latter are scattered off of
a bend. We have found that for soft potentials, the bend operates as
a potential barrier  allowing transmission for supercritical and reflection
for subcritical values of the localized excitation's initial speed.
However, there can also be narrow intervals of trapping (not captured
by the potential barrier picture). On the other hand, for hard
potentials, a more complex and partly initial condition
(i.e., kinetic energy) dependent picture
emerges. For small bend strengths, the inhomogeneity acts as a potential
well, but for higher ones it can lead to trapping and even to reflection.

In this numerical study, we have mainly aimed at presenting the relevant
phenomenology, classifying it for Klein-Gordon chains with different
potentials, and providing a qualitative explanation of the relevant
findings. It would naturally be of interest to attempt to obtain more
quantitative, theoretically predicted, estimates for the transitions
observed herein. A variational viewpoint may prove to be useful in this
context.

\begin{acknowledgments}

The support of NSF-DMS-0204585, a University of Massachusetts
Faculty Research Grant, and of the Eppley Foundation for Research
is gratefully acknowledged (PGK). JC acknowledges an FPDI grant
from `La Junta de Andaluc\'{\i}a' and partial support under the
European Commission RTN project LOCNET, HPRN-CT-1999-00163.

\end{acknowledgments}

\newcommand{\noopsort}[1]{} \newcommand{\printfirst}[2]{#1}
  \newcommand{\singleletter}[1]{#1} \newcommand{\switchargs}[2]{#2#1}

\end{document}